\documentclass{elsartL}
\usepackage{graphicx}
\usepackage{amssymb}
\usepackage{amsmath}
\usepackage{physics}
\usepackage{mathrsfs}
\usepackage{mathtools}
\usepackage{rotating}
\usepackage{appendix}
\newcommand{\RNum}[1]{\uppercase\expandafter{\romannumeral #1\relax}}
\newcommand{\avg}[1]{\left< #1 \right>} 
\begin{document}

\begin{frontmatter}
\title{Extreme-value Statistics: Rudiments and applications}
\author{Evangelos Matsinos}

\begin{abstract}This study provides a summary of the theory which enables the analysis of extreme values, i.e., of measurements acquired from the observation of extraordinary/rare physical phenomena. The 
formalism is developed in a transparent way, tailored to the application to real-world problems. Three examples of the application of the theory are detailed: first, an old problem, relating to the return 
period of floods caused by the Rh{\^o}ne river, is revisited; second, the frequency of occurrence and the severity of meteorite impacts on the Earth are examined; the data, which are analysed in the third 
example, relate to the longevity of supercentenarians.\\
\noindent {\it PACS 2010: 02.50.-r, 02.70.Rr, 02.70.Uu, 92.40.qp, 96.12.ke, 96.15.Qr, 96.20.Ka}
\end{abstract}
\end{frontmatter}

\section{\label{sec:Introduction}Introduction}

On 1 February 1953, the residents of the coastal regions in the Low Countries and in the United Kingdom (UK) woke up to the consequences of extensive flooding, caused by a storm tide in the North Sea, rising 
above the mean sea level to a crest of over $5.5$m. The death toll rose to $1\,836$ in the Netherlands, $326$ in the UK, and $28$ in Belgium \cite{NorthSeaFlood1953}; in addition, $230$ seamen perished in the 
North Sea. Hundreds of thousands of animals drowned, tens of thousands of houses and farms were inundated, ten times more houses were severely damaged.

For the Netherlands, a densely-populated country with two-thirds of its area vulnerable to flooding, the 1953 calamity was neither the first nor the worst in terms of fatalities and destruction of infrastructure. 
Since 838 CE, a total of $27$ floods have been documented \cite{FloodsInTheNetherlands}, $12$ of which had resulted in fatalities: $2\,437$ in 838; thousands in 1014; $60\,000$ in 1212; $36\,000$ in 1219; over 
$50\,000$ in 1287 (St.~Lucia's flood); over $25\,000$ in 1362 (St.~Marcellus's flood); between $2\,000$ and $10\,000$ in 1421 (St.~Elizabeth's flood); over $100\,000$ in 1530 (St.~Felix's flood); $20\,000$ in 
1570 (All Saints' flood); between $8\,000$ and $15\,000$ in 1703 (Great storm of 1703); $14\,000$ in 1717 (Christmas flood of 1717); and about $800$ in 1825 (February flood of 1825).

It therefore appears that, on average, the Netherlands is hit by one major storm tide per century. The country is currently protected against storm surges from the Atlantic via a complex system of dykes, dams, 
and floodgates, reinforced or constructed after the 1953 flood in accordance with predictions obtained by means of the Extreme-Value Theory (EVT). In essence, the Netherlands became the domain of application 
of the theory which purports at providing reliable estimates for the likelihood of occurrence of extraordinary/rare events, i.e., of events which (in the context of Taleb's book `The Black Swan' \cite{Taleb2010}) 
are considered ``extraordinary'' in the fictitious country of ``Mediocristan'' (where many of us \emph{believe} we live), yet ``ordinary'' in the country of ``Extremistan'' (where we \emph{do} live). Taleb 
argues that our bewildered reaction to extraordinary events may be traced to the fact that (as a rule) our analysts base their assessment of likelihood on the Gaussian (normal) distribution~\footnote{In an 
interesting 2012 short report \cite{Matthews2012}, Robert Matthews commented on an unexpected crisis in Goldman Sachs five years earlier, writing: ``David Viniar, CFO of Goldman Sachs, admits his perplexity 
after announcing that the loss of $27$~\% of the value of one of the firm's flagship funds represented a $25 \sigma$ effect. Such an event would not normally be expected to have occurred even once since the 
birth of the Universe.''}, even when the stochastic variable is not expected to follow it (as, for instance, the case is when considering the distribution of the largest value of normally-distributed 
data \cite{Bortkiewicz1922,Mises1923}).

One may argue that the seeds of the EVT were sown over $300$ years ago, with the dissertation \cite{Bernoulli1709} of the Swiss mathematician Nicolaus (Niclaus) Bernoulli (1687-1759), also known as Nicolaus I 
Bernoulli. At the end of p.~35 of his 1860 book `Die Mathematiker Bernoulli' \cite{Merian1860}, Peter Merian wrote: ``Im Juni 1709 promovirte Niclaus Bernoulli als \emph{Licentiatus juris}. Seine Dissertation: 
\emph{De usu Artis conjectandi in Jure}, welche die Todeserkl{\"a}rung Verschollener, Leibrenten und {\"a}hnliche Gegenst{\"a}nde behandelt, ist ein sehr bemerkenswerther Beitrag zur Wahrscheinlichkeitsrechnung. 
Leibniz war höchlich dar{\"u}ber erfreut und begl{\"u}ckw{\"u}nschte Johann Bernoulli {\"u}ber dieses neue Talent in seiner Familie; in der That, mit dem Auftreten von Niclaus Bernoulli fieng die Meinung an 
sich festzustellen, dass mathematische Begabung dem Namen Bernoulli angeboren sei.'' (In June 1709, Niclaus Bernoulli received his doctorate as \emph{Licentiate of Law}. His dissertation: \emph{On the Use of 
the Art of Conjecture in Law}, which deals with the declaration of death of missing persons, annuity incomes, and similar matters, is a very remarkable contribution to the probability theory. Leibniz was very 
pleased and complimented Johann Bernoulli on this new talent in his [i.e., in the Bernoulli] family; in fact, with the appearance of Niclaus Bernoulli, the opinion began to congeal that mathematical talent was 
inherent to the name Bernoulli.)

As the case sometimes is, the EVT had to wait for at least two centuries after Bernoulli's dissertation to reach adulthood and, even longer, to be applied to real-world problems. However, the applications of 
the EVT to numerous challenging situations have been rising at an accelerated pace over the past few decades. For instance, the technique has been used for providing solutions to problems in Civil Engineering, 
ranging from the safety of structures to Geology and Hydrology. Aviation engineers have applied the EVT to problems pertinent to the design of aircraft, aiming at improving performance and safeguarding against 
severe weather phenomena. The `prediction' of earthquakes in Seismology is another domain of application of the EVT. Regarding the climate crisis, one of the major worries of humankind at this time, the EVT 
has been used for predicting severe weather phenomena, e.g., droughts, rainfall, flooding, etc., and to make for our defences in time. Statistical studies, dealing with the detection of outliers in datasets, 
frequently used elements of the EVT. The application of the EVT to human physical performance has also been established, from the pioneering studies on old age by Emil Julius Gumbel (1891-1966) in the 1930s 
to the extreme performance of athletes \cite{Matthews2012}. Last but not least, the EVT has been applied (albeit with considerable delay) to subjects in the financial sector: for instance, to assess the 
probability of bank crises and market crashes, and to mitigate their consequences; regarding this last subject, Taleb has a plentiful supply of `juicy' details to relate \cite{Taleb2010}.

A number of books provide invaluable insight into the EVT, both in terms of the development of the concepts as well as of the mathematical formalism. I recommend the (chronologically-arranged) list of 
Refs.~\cite{Gumbel1958,Kinnison1983,Leadbetter1983,Coles2001,Sornette2006}. For those who would rather opt for a comprehensive discussion on why extraordinary events might not be so extraordinary as most of 
us think, without delving into the mathematics relevant to the subject, I recommend Taleb's book `The Black Swan' \cite{Taleb2010}, the only book I know of, which has been written by a mathematician and - to 
the rejoice of most (and dismay of the rest) - contains no mathematical formula.

A number of acronyms will be used in this study.
\begin{itemize}
\item PDF will stand for `Probability Density Function' (for continuous stochastic variables);
\item PMF will stand for `Probability Mass Function' (for discrete stochastic variables);
\item CDF will stand for `Cumulative Distribution Function';
\item PCC will stand for `Pearson Correlation Coefficient';
\item CL will stand for `confidence level' and CI for `confidence interval';
\item DB will stand for `database';
\item DoF will stand for `degree of freedom';
\item $\mathbb{Z}$ will stand for the set of all integers: $\mathbb{Z}_{\geq 0}$ will be the set of the non-negative integers, $\mathbb{Z}_{> 0}$ will be the set of the positive integers, etc.;
\item $\mathbb{N}^+_k$ will stand for the set of the integers between $1$ and $k \in \mathbb{Z}_{\geq 1}$, whereas $\mathbb{N}^0_k$ will stand for the set of the integers between $0$ and $k \in \mathbb{Z}_{\geq 0}$;
\item $\Delta t$ will denote the temporal interval associated with the extraction of the extreme values in each problem: depending on how these values are obtained, $\Delta t$ will be identified with the 
temporal interval between successive observations (measurements) of a physical system or with the bin width of the histogram which yielded the extreme values; and
\item Myr and Gyr will stand for million years ($10^6$ yr) and billion years ($10^9$ yr), respectively.
\end{itemize}

The aim of this study is to provide information to scientists, who do not have (but want to obtain) an overview of the EVT, both in terms of the theoretical background as well as of its application. The 
structure of this paper is as follows. The rudiments of the EVT are given in Section \ref{sec:Theory}, which is split into two parts: the first part relates to the number of the exceedances, a non-parametric, 
model-independent (i.e., distribution-free) method of analysis, featuring Order Statistics; the second part addresses the magnitude of the exceedances. The three classes of functions, which the extreme values 
asymptotically follow, are given in Section \ref{sec:MEVGeneralFormalism} and their properties are discussed further in Section \ref{sec:MEVProperties}; these two sections are fundamental to analysing the 
magnitude of the extreme values as well as to extracting reliable predictions of their frequency of occurrence. Section \ref{sec:MEVAnalysis} summarises the steps, which need to be followed in the general 
analysis procedure of the extreme values and in the extraction of predictions. Section \ref{sec:Applications} details the results of the application of the EVT to three selected real-world problems: to the 
floods caused by the Rh{\^o}ne river, a subject which Gumbel addressed in the first ever application of the EVT to a real-world problem in 1941; to the impacts of meteorites on the Earth; and to the extreme 
human longevity. Finally, a summary of the findings of this work is given in Section \ref{sec:Conclusions}.

\section{\label{sec:Theory}Theoretical considerations}

In much of the literature, the theory, which is used in the analysis of the extreme values acquired from observations of a physical phenomenon, is divided into two parts, one involving the number of 
occurrences of the extraordinary events, the other their magnitude. This practice gives the fallacious impression to the reader that there are two categories of extreme-value problems: those associated with 
the number of occurrences of the extraordinary events and the ones which are associated with their magnitude. In reality however, it is not the extreme-value problems which need to be categorised, but the 
methods of analysis of such problems.
\begin{itemize}
\item The first method of analysis, which was developed earlier, is non-parametric and relies on Order Statistics. In essence, the method transforms $n$ measurements $x_i$, $i \in \mathbb{N}^+_n$, of a physical 
quantity at the ratio level into measurements at the ordinal level. I shall elaborate on this issue in Section \ref{sec:MEVAnalysis}. At this point, it suffices to say that this method makes use only of the 
rank of the input data after they are arranged in ascending ($x_1 \leq x_2 \leq \dots \leq x_{n-1} \leq x_n$) or descending ($x_1 \geq x_2 \geq \dots \geq x_{n-1} \geq x_n$) order; the underlying distribution, 
from which the original data had been drawn, is of no relevance.
\item The second method of analysis accounts for the magnitude of the extraordinary events on the basis of a few (two or three) parameters entering three classes of functions.
\end{itemize}
In both cases, the same question is ultimately asked: how frequently are the extraordinary events expected to occur in future observations of the particular physical phenomenon? In my opinion, the estimates 
obtained with the second method are reliable, whereas those obtained with the first are approximate.

To scientists who have not yet come across the EVT, I recommend Kinnison's 1983 monograph \cite{Kinnison1983} as a good starting point. From my standpoint, the book retains a good balance between the introduction 
of the important concepts and the development of the mathematical formalism. Its main weakness bears on the overall shabbiness of the presentation: the book contains numerous orthographical, syntactical, 
typographical, and (even worse) computational mistakes; evidently, corrections to the text, which was typed on a typewriter, were not deemed worth the effort. I believe that this book deserves better: first, 
the material (text and all figures) must be updated so that it reflects the present-day expectations (owing to the remarkable advances in Computational Science over the past few decades, many of the phrases 
in the book need to be strongly revised or even removed~\footnote{On p.~10-6 of the book, Kinnison starts one hilarious (by today's standards) sentence with the condition: ``If a computer is available, \dots''}); 
second, it must be retyped using a modern word processor (preferably LaTeX, which would also enable the appearance of its mathematical relations in a form matching today's quality standards).

\subsection{\label{sec:DistributionNumberOfExceedances}Distribution of the number of the exceedances}

Kinnison sets out to develop the mathematical formalism, relevant to the PMF of the number of the exceedances, by introducing two datasets, the first one relating to historical measurements (reference dataset), 
the second to new (future) ones. In essence, Kinnison lays out the strategy, followed since the early days of development of the EVT, towards the establishment of a non-parametric, model-independent method of 
analysis, featuring Order Statistics: the analysis of a new dataset is not based on information \emph{extracted} from the historical measurements (e.g., on probabilities emerging thereof), but on the direct 
comparison of the two datasets (reference and new).

The development of the formalism rests upon two assumptions; these assumptions also apply to the development of the formalism in Section \ref{sec:MagnitudeOfExtremeValues}.
\begin{itemize}
\item All measurements have been drawn from the same distribution (regardless of which that distribution might be).
\item All measurements are statistically independent (uncorrelated).
\end{itemize}
The first assumption necessitates the absence of time-dependent effects in the sampling of all values. When it comes down to measurements of physical quantities, particular attention must be paid to the 
fulfilment of the second assumption. Kinnison stresses the importance of this condition with one good example (see Ref.~\cite{Kinnison1983}, see p.~2-2): ``Daily maxima of pollutant concentrations are dependent 
(correlated) because the causes of pollution are phenomena which last for many days.''

One quantity, which provides a rough, yet model-independent estimate for the frequency of occurrence of extraordinary events, is the `empirical return period' $T_e$. Provided that the historical measurements 
$\mathfrak{X}_i, i \in \mathbb{N}^+_n$, have been arranged in ascending order, $T_e$ - a function of $\mathfrak{X}_i$ - is defined as
\begin{equation} \label{eq:EQ000}
T_e (\mathfrak{X}_i) = \frac{n+1}{n+1-i} \, \, \, .
\end{equation}
In fact, the general expression for $T_e (\mathfrak{X}_i)$ involves the normalised-rank array $r_i$, which will be defined later on, in Eq.~(\ref{eq:EQ031_5}):
\begin{equation} \label{eq:EQ000_5}
T_e (\mathfrak{X}_i) = \frac{1}{1-r_i} \, \, \, .
\end{equation}
The choice $A=0$ in Eq.~(\ref{eq:EQ031_5}), which is generally favoured by Kinnison, leads to Eq.~(\ref{eq:EQ000}). I mention this quantity for the sake of completeness; a more reliable estimate for the 
return period of extraordinary events will be introduced in Section \ref{sec:MEVGeneralFormalism}.

\subsubsection{\label{sec:DNEGeneralFormalism}General formalism}

Let $n$ and $N$ denote the dimensions of the reference dataset (arranged in descending order) and of the new dataset (similarly arranged), respectively. Regarding $n$ and $N$, no assumption needs to be made. 
Kinnison next raises the question: What is the distribution of the stochastic variable $x$, if $x$ represents the number of elements in the new dataset which are at least equal to (i.e., equal to or larger 
than) the $m$-th element of the reference dataset? Evidently, this distribution depends on $m$.

Let the PMF of the number of the exceedances $x$ be denoted by $w(x;n,m,N)$, where $x \in \mathbb{N}^0_N$ and $m \in \mathbb{N}^+_n$. Regardless of the type of the (common) distribution from which the two 
samples have been drawn, the following relation emerges.
\begin{align} \label{eq:EQ001}
w(x;n,m,N) &= \frac{C(m-1+x,m-1) \, C(n-m+N-x,n-m)}{C(n+N,n)}\nonumber\\
 &\equiv \frac{m}{n+N} \, \frac{C(n,m) \, C(N,x)}{C(n+N-1,m-1+x)} \, \, \, ,
\end{align}
where $C(a,b)$, with $a, b \in \mathbb{Z}_{\geq 0}$ and $a \geq b$, is the binomial coefficient
\begin{equation} \label{eq:EQ002}
C(a,b) \coloneqq \binom{a}{b} = \frac{a!}{(a-b)! \, b!} \, \, \, .
\end{equation}
As all binomial coefficients are \emph{positive}, $w(x;n,m,N)>0$; Kolmogorov's first axiom (non-negative probability of the events belonging to the event space) is fulfilled. Using the identity
\begin{equation} \label{eq:EQ003}
\sum_{x=0}^{a} C(a+b-x,b) \, C(c+x,c) = C(a+b+c+1,b+c+1) \, \, \, ,
\end{equation}
which is valid $\forall a, b, c \in \mathbb{Z}_{\geq 0}$, one obtains (for $a=N$, $b=n-m$, and $c=m-1$)
\begin{equation} \label{eq:EQ004}
\sum_{x=0}^{N} w(x;n,m,N) = 1 \, \, \, .
\end{equation}
As a result, Kolmogorov's second axiom (normalisation condition: the probability of the entire event space is equal to $1$) is also fulfilled, hence the quantity $w(x;n,m,N)$ of Eq.~(\ref{eq:EQ001}) is 
indeed~\footnote{In this case, Kolmogorov's third axiom is fulfilled because any two events, characterised by $x=x_1$ and $x=x_2$ with $x_1, x_2 \in \mathbb{N}^0_N$ and $x_1 \neq x_2$, cannot occur 
simultaneously; hence the probability that either of them occurs is equal to the sum of the probabilities corresponding to their separate occurrence.} a PMF.

\begin{figure}
\begin{center}
\includegraphics [width=15.5cm] {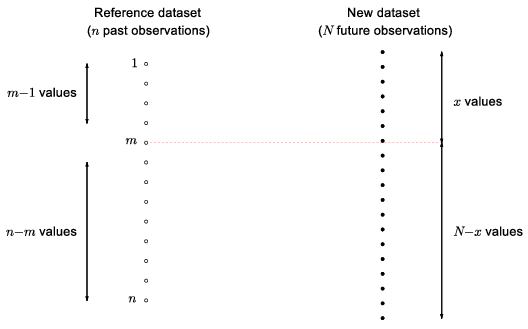}
\caption{\label{fig:Sets}The dataset on the left corresponds to the historical measurements (reference dataset); the one on the right to the new measurements (new dataset). Both sets are assumed arranged in 
descending order. Regarding the dimensions $n$ and $N$, no assumption needs to be made. The figure facilitates the extraction of the PMF of the number of the exceedances $x$ of Eq.~(\ref{eq:EQ001}).}
\vspace{0.5cm}
\end{center}
\end{figure}

Parenthetically, the PMF of Eq.~(\ref{eq:EQ001}) is not difficult to obtain, see Fig.~\ref{fig:Sets}. As $m-1$ elements in the ordered reference dataset are at least equal to the $m$-th value, there are 
$n-m$ elements below the $m$-th value. In the new dataset, $x$ elements are at least equal to the $m$-th value in the ordered reference dataset, whereas $N-x$ elements correspond to smaller values. Evidently, 
there are $C(m-1+x,m-1)$ ways of selecting $m-1$ out of $m-1+x$ objects, $C(n-m+N-x,n-m)$ ways of selecting $n-m$ out of $n-m+N-x$ objects, and $C(n+N,n)$ ways of selecting $n$ out of $n+N$ objects. One thus 
obtains Eq.~(\ref{eq:EQ001}).

The corresponding CDF will be denoted by $W(x;n,m,N)$; of course,
\begin{equation} \label{eq:EQ005}
W(x;n,m,N) \coloneqq \sum_{i=0}^{x} w(i;n,m,N) \, \, \, .
\end{equation}

Analytical expressions for the first two moments of $w(x;n,m,N)$ will next be obtained.
\begin{align} \label{eq:EQ006}
\mu_1 \coloneqq \avg{x} &= \sum_{x=0}^{N} x \, w(x;n,m,N)\nonumber\\
 &= m \left( C(n+N,n) \right)^{-1} \sum_{x=0}^{N-1} C(m+x,m) \, C(n-m+N-1-x,n-m)\nonumber\\
 &= m \, C(n+N,n+1) / C(n+N,n)\nonumber\\
 &= \frac{m N}{n+1} \, \, \, ,
\end{align}
where use has been made of Eq.~(\ref{eq:EQ003}) (with $a=N-1$, $b=n-m$, and $c=m$). Finally,
\begin{align} \label{eq:EQ007}
\mu^\prime_2 &= \sum_{x=0}^{N} x^2 \, w(x;n,m,N)\nonumber\\
 &= m \left( C(n+N,n) \right)^{-1} \sum_{x=0}^{N-1} (x+1) \, C(m+x,m) \, C(n-m+N-1-x,n-m)\nonumber\\
 &= m \left( C(n+N,n) \right)^{-1} \Big((m+1) \sum_{x=0}^{N-2} C(m+1+x,m+1) \, C(n-m+N-2-x,n-m)\nonumber\\
 &\qquad\qquad\qquad\qquad\qquad+ \sum_{x=0}^{N-1} C(m+x,m) \, C(n-m+N-1-x,n-m) \Big)\nonumber\\
 &= m \big( (m+1) C(n+N,n+2) + C(n+N,n+1) \big) / C(n+N,n)\nonumber\\
 &= \frac{m N}{n+1} \left( \frac{(m+1) (N-1)}{n+2} + 1 \right) \, \, \, ,
\end{align}
resulting in the central moment
\begin{equation} \label{eq:EQ008}
\mu_2 \coloneqq {\rm Var} (x) = \frac{m N (n-m+1) (n+N+1)}{(n+1)^2 (n+2)} \, \, \, .
\end{equation}
Evidently, ${\rm Var} (x) \sim N^2$ for large $N$ and ${\rm Var} (x) \sim n^{-1}$ for large $n$. Put into words, Eq.~(\ref{eq:EQ008}) suggests that the variance increases quadratically with $N$ and decreases 
with $n$. Last but not least, $\mu_2$ attains its maximum for $m=(n+1)/2$, i.e., for the median of the measurements (an element of the set in case of odd $n$) comprising the reference dataset.

One of the frequently-asked questions in the extraction of predictions concerns the probability $p$ that the largest value of the reference set (i.e., $m=1$) will be at least equalled (at least once) in a 
set of $N$ future measurements. Evidently, the probability that the largest value will \emph{not} be at least equalled is given by
\begin{equation} \label{eq:EQ009}
w(0;n,1,N) = \frac{n}{n+N} \, \, \, ,
\end{equation}
implying that $p=1-w(0;n,1,N)=N/(n+N)$. For the sake of example, if the reference set comprises $n=100$ measurements, then the probability that its largest value \emph{will} be at least equalled in a set of 
$N=900$ new measurements is $90$~\%. The predicament, wherein the analysts find themselves (when extracting predictions from historical measurements), bears on the fact that the number of historical 
measurements is frequently smaller than what would be necessitated for obtaining \emph{reliable} predictions. For instance, to be able to support the claim, with some degree of confidence (say, $90$~\%), 
that the largest earthquake of a reference dataset will not be at least equalled over the subsequent $k$ decades, the historical measurements must span (on average) a temporal interval of $9 k$ decades.

Kinnison finalises the main part of Chapter 6 of his book with the asymptotic behaviour of the PMF of Eq.~(\ref{eq:EQ001}) in two interesting (for applications) limiting cases. Such approximations were more 
useful in the days when affordable computers were unavailable; although they may provide some insight, they tend to be rather superfluous today.

\subsubsection{\label{sec:DNEAsymptotic1}Asymptotic behaviour for large $N$ and small $n$}

In this case, $m$ is also `small'. Kinnison introduces the variable $q \coloneqq x/N$, which (given that, for all intents and purposes, $N \to \infty$) he treats as `continuous'. In that case, the PMF of 
Eq.~(\ref{eq:EQ001}) is approximated by the PDF
\begin{equation} \label{eq:EQ010}
f(q;n,m) = m C(n,m) \, q^{m-1} \, (1-q)^{n-m} \, \, \, .
\end{equation}
As a result, the CDF of Eq.~(\ref{eq:EQ005}) takes the form
\begin{align} \label{eq:EQ011}
F(q;n,m) &= \int_{0}^{q} f(t;n,m) dt = m C(n,m) \int_{0}^{q} t^{m-1} (1-t)^{n-m} dt\nonumber\\
 &\equiv m C(n,m) \, B(q;m,n-m+1) \, \, \, ,
\end{align}
where
\begin{equation} \label{eq:EQ012}
B(q;a,b) \coloneqq \int_{0}^{q} t^{a-1} (1-t)^{b-1} dt
\end{equation}
denotes the incomplete beta function, i.e., the product of the regularised incomplete beta function $I_q (a,b)$ and the beta function 
\begin{equation} \label{eq:EQ013}
B(a,b) = \frac{\Gamma(a) \Gamma(b)}{\Gamma(a+b)} \, \, \, ,
\end{equation}
where $\Gamma$ is the gamma function. Algorithms enabling the evaluation of the incomplete beta function are available, e.g., see \cite{Press2007}, Chapter 6.4, pp.~272-273; it should be mentioned that what 
Ref.~\cite{Press2007} calls incomplete beta function is actually the regularised incomplete beta function $I_q (a,b)$. The function BETA.DIST in Microsoft Excel can be used for evaluating the integral of 
Eq.~(\ref{eq:EQ012}): BETA.DIST$(q,a,b,1)$ returns the ratio $B(q;a,b)/B(a,b)$.

\subsubsection{\label{sec:DNEAsymptotic2}Asymptotic behaviour for large $N$ and $n$}

Kinnison addresses two cases of interest:
\begin{itemize}
\item the $m$-th element is close to the median of the $n$ measurements and the ratio $m/n$ is a near constant with increasing $n$, and
\item $m \ll n$.
\end{itemize}

In the former case, the distribution of the number of the exceedances approaches (asymptotically) the normal distribution, see Ref.~\cite{Gumbel1958} Section 2.2.6. This distribution is also known as the 
`distribution of normal exceedances'.

In the latter case, which is known as the `law of rare exceedances',
\begin{equation} \label{eq:EQ014}
w(x;n,m,N) \approx C(m-1+x,m-1) \frac{n^m N^x}{(n+N)^{m+x}} \, \, \, .
\end{equation}
Consequently, the probability that the $m$-th value of the reference dataset will not be at least equalled ($x=0$) in the new dataset is approximated by
\begin{equation} \label{eq:EQ015}
w(0;n,m,N) \approx \left( \frac{n}{n+N} \right)^m \, \, \, .
\end{equation}
Finally, the probability that the largest value of the reference dataset will be at least equalled exactly $x$ times is approximated by
\begin{equation} \label{eq:EQ016}
w(x;n,1,N) \approx \frac{n}{n+N} \left( \frac{N}{n+N} \right)^x \, \, \, ,
\end{equation}
which is a geometrical progression with common ratio $N/(n+N)<1$. The mean and the variance of the number of rare exceedances are obtained from Eqs.~(\ref{eq:EQ006},\ref{eq:EQ008}): $\avg{x} \approx m$ and 
${\rm Var} (x) \approx 2m$.

\subsection{\label{sec:MagnitudeOfExtremeValues}Magnitude of the extreme values}

If I could choose \emph{two} founders of Modern Statistics, I would have selected Karl Pearson (1857-1936), whose academical advisor was Francis Galton (1822-1911), and Ronald Fisher (1890-1962). Had Alfred 
Nobel's will prescribed that the Prize could also be awarded in Mathematics, Pearson and Fisher would undoubtedly have received it. Both were fortunate to watch Statistics gradually moving away from its 
$18$-th century `marginal status' (``when wealthy gamblers called upon mathematicians to determine the correct odds in their games, so they could find the best betting strategies'' \cite{Kinnison1983}, 
p.~3-9) and becoming an integral branch of Mathematics.

\subsubsection{\label{sec:MEVGeneralFormalism}General formalism}

In their seminal 1928 paper \cite{Fisher1928}, Fisher and Leonard Henry Caleb Tippett (1902-1985) examined the CDF $\Psi$ of the largest values in datasets drawn from an arbitrary PDF $f$, and concluded that - 
regardless of the type of the distribution $f$ - the distribution $\Psi$ approaches one of three possible classes of functions~\footnote{This result is known as Fisher-Tippett or Fisher-Tippett-Gnedenko 
theorem.}, as the dimension $n$ of the sample increases indefinitely.
\begin{equation} \label{eq:EQ017}
\Psi(y) = \left\{
\begin{array}{rl}
e^{-e^{-y}} & \text{Gumbel class, denoted herein as $\Psi_{\rm \RNum{1}}(y)$}\\
e^{-y^{-\zeta}} & \text{Fr{\'e}chet class, denoted herein as $\Psi_{\rm \RNum{2}}(y)$}\\
e^{-(-y)^{-\zeta}} & \text{Weibull class, denoted herein as $\Psi_{\rm \RNum{3}}(y)$}\\
\end{array} \right.
\end{equation}
The support is different in the three cases: in the Gumbel class, $y \in \mathbb{R}$; in the Fr{\'e}chet class, $y \in \mathbb{R}_{\geq 0}$ (and $\zeta \in \mathbb{R}_{> 0}$); in the Weibull class, 
$y \in \mathbb{R}_{\leq 0}$ (and $\zeta \in \mathbb{R}_{< 0}$). Unbeknownst to Fisher and Tippett at the time of their publication, the second class, $\Psi_{\rm \RNum{2}}(y)$, had already been discovered by 
Ren{\'e} Maurice Fr{\'e}chet (1878-1973) \cite{Frechet1927}.

Due to two reasons, the Gumbel class is the most frequently employed in the extreme-value analyses. First, the extreme values in samples, drawn from a wide family of PDFs, follow $\Psi_{\rm \RNum{1}}(y)$; 
this does not only apply to the normal distribution (e.g., see Appendix \ref{App:AppA}), but also to other important distributions (e.g., exponential, gamma, and $\chi^2$). Second, the Gumbel class enables 
the description of the data on the basis of just two parameters. The Fr{\'e}chet class is primarily associated with the Lorentzian (Cauchy) distribution. Finally, the Weibull class, which is frequently 
employed in survival analysis, in failure analysis, and in weather forecasting, is associated with PDFs which are bounded, e.g., with the Weibull and the beta distributions.

The three distributions of Eqs.~(\ref{eq:EQ017}) are related via logarithmic transformations. For the sake of example, if the distribution of a variable $y$ belongs to the Fr{\'e}chet class, then the 
corresponding distribution of $\ln y$ belongs to the Gumbel class, see Ref.~\cite{Kinnison1983}, p.~3-9.

\begin{figure}
\begin{center}
\includegraphics [width=15.5cm] {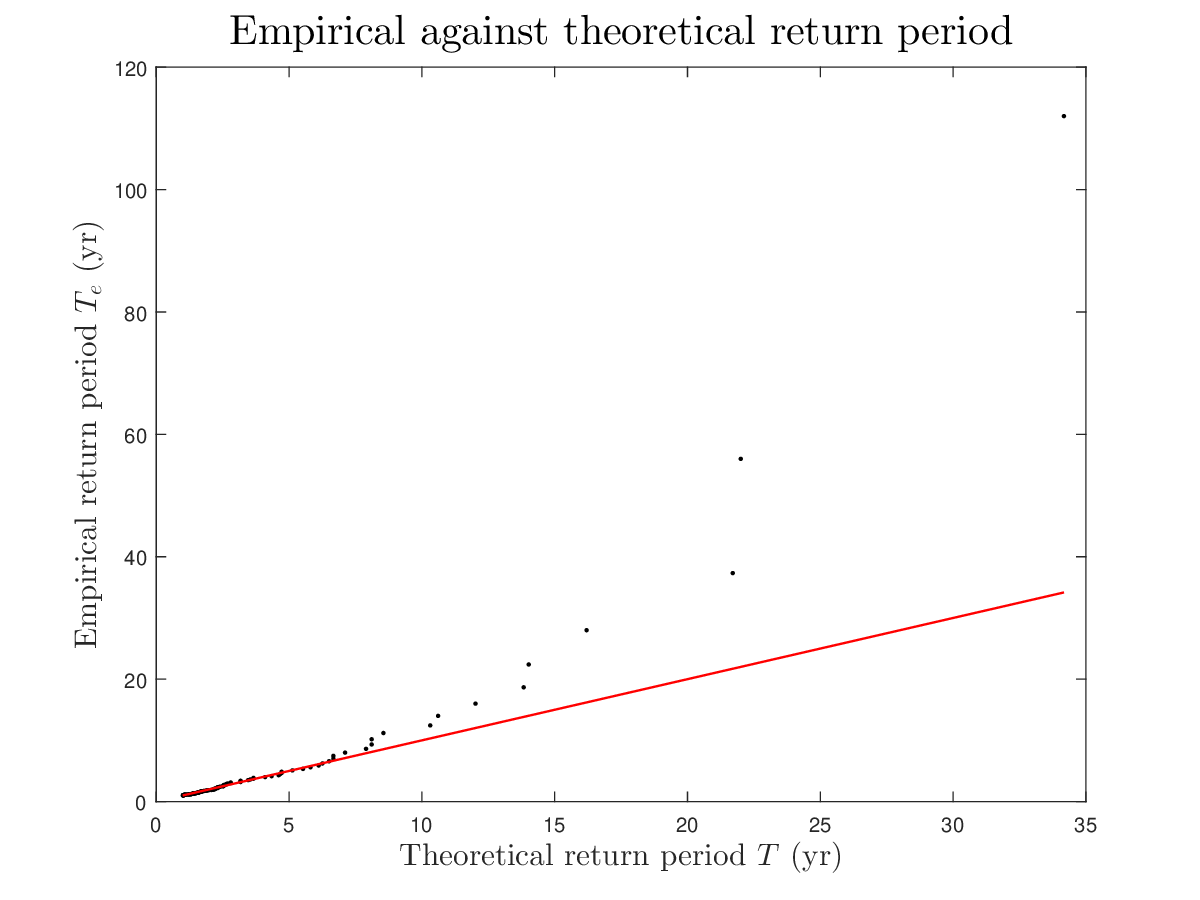}
\caption{\label{fig:ComparisonReturnPeriods}The empirical return period $T_e$ of Eq.~(\ref{eq:EQ000}) against the theoretical return period of Eq.~(\ref{eq:EQ017_4}) for the data of Section \ref{sec:Floods}. 
The red solid straight line represents the equation $T_e=T$.}
\vspace{0.5cm}
\end{center}
\end{figure}

One question when dealing with extraordinary events pertains to the frequency of their occurrence. To this end, the return period, the average temporal separation between successive occurrences of such 
events, is defined according to the formula
\begin{equation} \label{eq:EQ017_4}
T (y_t) = (1-\Psi(y_t))^{-1} \, \, \, ,
\end{equation}
where $y_t$ represents the threshold value (lower bound) of the extraordinary events whose return period is sought. By definition, the return period $T (y_t)$ is nothing but the average number of new 
observations~\footnote{To transform the quantities $T_e (\mathfrak{X}_i)$ and $T (y_t)$ of Eqs.~(\ref{eq:EQ000},\ref{eq:EQ017_4}) into `times', one needs to multiply both quantities by $\Delta t$, as defined 
at the end of Section \ref{sec:Introduction}.} required until the appearance of one value $y \geq y_t$. The return period $T (y_t)$ is frequently referred to as `theoretical'. Unlike the empirical return 
period $T_e$ of Eq.~(\ref{eq:EQ000}), which is evaluated using the \emph{rank} of the historical measurement $\mathfrak{X}_i$, not $\mathfrak{X}_i$ itself, $T (y_t)$ involves the probability 
$P (y \geq y_t) = 1 - \Psi(y_t)$, hence it is the correct measure of the frequency of occurrence of values $y \geq y_t$. The two return periods are shown in Fig.~\ref{fig:ComparisonReturnPeriods} against one 
another for the data analysed in Section \ref{sec:Floods}; in that case, Eq.~(\ref{eq:EQ000}) significantly overestimates the return period at the high end of the extreme values, suggesting (in such cases) a 
sizeable underestimation of the risk if the risk assessment is based on that equation.

Before continuing, let me clarify one issue which might appear perplexing to those who start studying the details of the EVT. Let $x_i, i \in \mathbb{N}^+_n$, denote independent and identically-distributed 
stochastic variables representing the results of $n$ observations. As dimensional analysis suggests (the quantity $y$ in Eqs.~(\ref{eq:EQ017}) is dimensionless, whereas $x$ usually carries units of 
measurement), $x$ cannot be identified with $y$. In extreme-value analyses, the variable $y$ is known as \emph{reduced} variable, and is associated with the actual measurements via the relation
\begin{equation} \label{eq:EQ017_5}
y = \frac{x-\mu}{\alpha} \, \, \, ,
\end{equation}
where $\mu \in \mathbb{R}$ is known as location parameter and $\alpha \in \mathbb{R}_{>0}$ is known as scale parameter; both these parameters have the physical dimension of the measurements $x$. Estimates 
for $\mu$ and $\alpha$ may be obtained directly (i.e., from the sample mean and variance) or indirectly (i.e., by means of an optimisation scheme). Equation (\ref{eq:EQ017_5}) is reminiscent of the relation 
between a quantity $x$ following the (general) normal distribution and the standard normal deviate $z$ (standardised residual).

It can be shown that the three classes of Eqs.~(\ref{eq:EQ017}) emerge from a general class, known as Generalised Extreme Value (GEV) distribution:
\begin{equation} \label{eq:EQ018}
\tilde{\Psi} (y) = e^{-(1 + \xi y)^{-\xi^{-1}}}
\end{equation}
for different values of the shape parameter $\xi$. For convenience, let me introduce the variable $\zeta \coloneqq \xi^{-1}$.

The case $\xi \to 0^\pm$ will be considered first, in which case $\zeta \to \pm \infty$ (the signs match). Under the aforementioned substitution,
\begin{equation} \label{eq:EQ019}
\tilde{\Psi} (y) = e^{-(1 + y/\zeta)^{-\zeta}} = e^{-f(y)} \, \, \, ,
\end{equation}
where (obviously) $f(y) \coloneqq (1 + y/\zeta)^{-\zeta}$. The Maclaurin expansion of the function $f(y)$ gives
\begin{equation} \label{eq:EQ020}
f(y) = f(0) + \frac{y}{1!} f^{(1)}(0) + \frac{y^2}{2!} f^{(2)}(0) + \frac{y^3}{3!} f^{(3)}(0) + \frac{y^4}{4!} f^{(4)}(0) + \dots \, \, \, ,
\end{equation}
where the first four derivatives of the function $f$ are given by
\begin{align} \label{eq:EQ021}
f^{(1)} (y) &= - \left( 1 + \frac{y}{\zeta} \right)^{-\zeta-1}\nonumber\\
f^{(2)} (y) &= \frac{\zeta+1}{\zeta} \left( 1 + \frac{y}{\zeta} \right)^{-\zeta-2}\nonumber\\
f^{(3)} (y) &= - \frac{(\zeta+1)(\zeta+2)}{\zeta^2} \left( 1 + \frac{y}{\zeta} \right)^{-\zeta-3}\nonumber\\
f^{(4)} (y) &= \frac{(\zeta+1)(\zeta+2)(\zeta+3)}{\zeta^3} \left( 1 + \frac{y}{\zeta} \right)^{-\zeta-4} \, \, \, .
\end{align}
For $\zeta \to \pm \infty$,
\begin{align} \label{eq:EQ022}
f^{(1)} (0) &= -1\nonumber\\
f^{(2)} (0) &= 1\nonumber\\
f^{(3)} (0) &= -1\nonumber\\
f^{(4)} (0) &= 1 \, \, \, .
\end{align}
Evidently, Eq.~(\ref{eq:EQ020}) takes the form
\begin{equation} \label{eq:EQ023}
f (y) = 1 - \frac{y}{1!} + \frac{y^2}{2!} - \frac{y^3}{3!} + \frac{y^4}{4!} - \dots = \sum_{k=0}^{\infty} \frac{(-y)^k}{k!} \equiv e^{-y}
\end{equation}
and $\tilde{\Psi} (y)$ of Eq.~(\ref{eq:EQ019}) reduces to the Gumbel class ($\Psi_{\rm \RNum{1}}(y)$).

When $\xi \neq 0$, $\zeta$ is finite and non-zero, and (of course) $\text{sgn} (\xi) = \text{sgn} (\zeta)$, where sgn denotes the sign function. Whichever the case, the argument $1 + y/\zeta$ cannot be 
negative. When $\zeta>0$, one obtains (after the replacement $1 + y/\zeta \to y$) the Fr{\'e}chet class ($\Psi_{\rm \RNum{2}}(y)$, for $y \in \mathbb{R}_{\geq 0}$). When $\zeta<0$, the exponent of argument 
$1 + y/\zeta$ is positive, and one obtains (after the replacement $1 + y/\zeta \to - y$) the Weibull class ($\Psi_{\rm \RNum{3}}(y)$, for $y \in \mathbb{R}_{\leq 0}$).

\subsubsection{\label{sec:MEVProperties}Some properties of the three classes of Eqs.~(\ref{eq:EQ017})}

The PDF $\psi_{\rm \RNum{1}}(y)$ corresponding to the Gumbel class can be obtained via differentiation of the CDF $\Psi_{\rm \RNum{1}}(y)$; the result is: $\psi_{\rm \RNum{1}}(y)=e^{-y-e^{-y}}$. The average 
of this distribution is obtained as follows.
\begin{equation} \label{eq:EQ024}
\avg{y} = \int_{-\infty}^{\infty} \, y \, e^{-y-e^{-y}} \, dy = - \int_{0}^{\infty} \, e^{-t} \, \ln t \, dt = \gamma \, \, \, ,
\end{equation}
where $\gamma \approx 0.577216$ is Euler's constant, and use has been made of Relation 4.331.1 of Ref.~\cite{Gradshteyn2007}, p.~571. Regarding the raw variance of this distribution, one obtains
\begin{equation} \label{eq:EQ025}
\avg{y^2} = \int_{-\infty}^{\infty} \, y^2 \, e^{-y-e^{-y}} \, dy = \int_{0}^{\infty} \, e^{-t} \, \left( \ln t \right)^2 \, dt = \frac{\pi^2}{6} + \gamma^2 \, \, \, ,
\end{equation}
where use has been made of Relation 4.335.1 of Ref.~\cite{Gradshteyn2007}, p.~572. Consequently,
\begin{equation} \label{eq:EQ025}
{\rm Var} (y) = \frac{\pi^2}{6} \, \, \, .
\end{equation}
Fisher and Tippett also give the third and fourth central moments of the Gumbel distribution \cite{Fisher1928}. Finally, the median of the distribution, obtained as the solution of the equation 
$\Psi_{\rm \RNum{1}}(y) = 1/2$, is equal to $- \ln \left( \ln 2 \right) \approx 0.366513$.

The PDF $\psi_{\rm \RNum{2}}(y)$ corresponding to the Fr{\'e}chet class can be obtained via differentiation of the CDF $\Psi_{\rm \RNum{2}}(y)$; the result is: $\psi_{\rm \RNum{2}}(y)=\zeta e^{-y^{-\zeta}} y^{-\zeta-1}$. 
The average of this distribution is given by the relation
\begin{equation} \label{eq:EQ026}
\avg{y} = \zeta \, \int_{0}^{\infty} \, y \, e^{-y^{-\zeta}} \, y^{-\zeta-1} \, dy = \Gamma (1-\zeta^{-1}) \, \, \, ,
\end{equation}
a result obtained after using the substitution $t=y^{-\zeta}$ and valid for $\zeta > 1$ (otherwise, $\avg{y}$ is infinite). Regarding the raw variance of this distribution, one obtains
\begin{equation} \label{eq:EQ027}
\avg{y^2} = \zeta \, \int_{0}^{\infty} \, y^2 \, e^{-y^{-\zeta}} \, y^{-\zeta-1} \, dy = \Gamma (1-2\zeta^{-1}) \, \, \, ,
\end{equation}
a result valid for $\zeta > 2$ (otherwise, $\avg{y^2}$ is infinite). Consequently, for $\zeta > 2$,
\begin{equation} \label{eq:EQ028}
{\rm Var} (y) = \Gamma (1-2\zeta^{-1}) - \left( \Gamma (1-\zeta^{-1}) \right)^2 \, \, \, .
\end{equation}
Finally, the median of the distribution is equal to $\left( \ln 2 \right)^{-\zeta^{-1}}$.

The PDF $\psi_{\rm \RNum{3}}(y)$ corresponding to the Weibull class can be obtained via differentiation of the CDF $\Psi_{\rm \RNum{3}}(y)$; the result is: $\psi_{\rm \RNum{3}}(y)=-\zeta e^{-(-y)^{-\zeta}} (-y)^{-\zeta-1}$. 
The average of this distribution is given by the relation
\begin{equation} \label{eq:EQ029}
\avg{y} = - \Gamma (1-\zeta^{-1}) \, \, \, ,
\end{equation}
a formula valid $\forall \zeta \in \mathbb{R}_{< 0}$. As to the raw variance of this distribution, one obtains
\begin{equation} \label{eq:EQ030}
\avg{y^2} = \Gamma (1-2\zeta^{-1}) \, \, \, ,
\end{equation}
a formula also valid $\forall \zeta \in \mathbb{R}_{< 0}$. Consequently,
\begin{equation} \label{eq:EQ031}
{\rm Var} (y) = \Gamma (1-2\zeta^{-1}) - \left( \Gamma (1-\zeta^{-1}) \right)^2 \, \, \, .
\end{equation}
Finally, the median of the distribution is equal to $- \left( \ln 2 \right)^{-\zeta^{-1}}$.

\subsubsection{\label{sec:MEVAnalysis}Real-world problems: Analysis of the magnitude of the extreme values}

On pp.~2-3 to 2-6 of his book, Kinnison defines the measurement as the ``activity of mapping or assigning numbers to objects or observations,'' and distinguishes among four levels (types) of measurement.
\begin{itemize}
\item For measurements at the \emph{nominal} (lowest) level, objects are simply identified (and counted). For instance, a measurement of this type may present `$30$ water-soluble coloured marker pens'.
\item For measurements at the \emph{ordinal} or \emph{ranking} level, objects are ordered. For instance, a measurement of this type may present `eighteen football teams in the Fu{\ss}ball-Bundesliga at the 
end of a year's championship'; the teams are ranked in accordance with their results during that year.
\item For measurements at the \emph{interval} level, numerical values are assigned to objects. For instance, one estimates the magnitude of an earthquake from its seismic moment (moment magnitude scale). One 
may then assess the approximate (on average, given the dependence of the effects on the type of earthquake and on the depth of its hypocenter) amount of damage due to specific earthquakes; for instance, an 
earthquake, categorised as $6 {\rm M_w}$, will release about the same energy as the atomic bomb dropped in Hiroshima, whereas one, categorised as $7 {\rm M_w}$, will release about $30$ times as much.
\item For measurements at the \emph{ratio} (highest) level, numerical values, including the value of (what I call a `meaningful') $0$, are assigned to objects. Kinnison defines the value of $0$ as ``the 
measure which defines the absence of a quantity.'' In practice, this implies that the ratio of the numerical values of two such measurements enables the straightforward comparison of the measured quantities, 
as (for instance) the case is with the length (about $29.8$m) and the width (about $19.2$m) of the Parthenon: by dividing the two values, one draws the conclusion that one side of the rectangular axial cross 
section of the temple is about $55$~\% larger than the other. Evidently, the straightforward comparison is not always possible~\footnote{Measurements at one specific level are always valid measurements at 
all lower levels; the reverse is not necessarily true.} for measurements at the interval level. For instance, the ratio of the values of the released energy in the two aforementioned earthquakes, categorised 
as $7 {\rm M_w}$ and $6 {\rm M_w}$, is not $7/6$, but about $30$. Measurements of the temperature, expressed in the Celsius and Fahrenheit scales, are also not measurements at the ratio level (as $0^\circ$C 
and $0^\circ$F do not signify ``absence of temperature'').
\end{itemize}

I shall next detail the general procedure which is followed in the analysis of the magnitude of the extreme values which are obtained from measurements of a physical quantity \emph{at the ratio level}. One 
starts with a (reference) dataset of $n$ such extreme values $x_i$. On the basis of an analysis of these values, one obtains estimates for the parameters $\mu$ and $\alpha$ of Eq.~(\ref{eq:EQ017_5}) (as well 
as for their uncertainties and correlation). This is all one needs in order that predictions relating to future events (e.g., to the return periods of extraordinary events) be obtained. Regarding the 
extraction of estimates for the parameters $\mu$ and $\alpha$, Kinnison mentions three options.

The first option involves ordinary linear regression on the datapoints of the probability plot~\footnote{This is essentially one of the forms of what is known today as $Q-Q$ plot, a graphical method for 
comparing two probability distributions by plotting their quantiles against each other: for the purposes of this work, the data quantiles are plotted against the `theoretical' quantiles ($k_i$, to be defined 
shortly).}.
\begin{itemize}
\item The extreme values $x_i$ of the reference dataset are arranged in ascending order.
\item The normalised-rank array (also known as the array of the cumulative relative frequencies) is constructed using the relation
\begin{equation} \label{eq:EQ031_5}
r_i = \frac{i-A}{n+1-2 A} \, \, \, ,
\end{equation}
yielding $\sum_{i=1}^n r_i = n/2$ $\forall A \in \mathbb{R}$. Various values of the parameter $A \in [0,1]$ have been employed in the literature for defining the plotting positions $r_i$. Although Kinnison 
favours $A=0$ (without providing justification), I frequently use $A=3/8$, the appropriate choice for the normal probability plot \cite{Blom1958}. When it comes down to fixing the value of the quantity $A$ 
in a problem, a meaningful approach might be to test the sensitivity of the PCC to the choice of $A$ (for the dataset in question), and select the $A$ value which maximises the PCC.
\item The array corresponding to the uniform order statistic medians $k_i \coloneqq \Psi^{-1} (r_i)$ is obtained $\forall i \in \mathbb{N}^+_n$, where $\Psi^{-1}$ denotes the inverse function (i.e., the quantile 
function) of (any of the three forms of) $\Psi$ of Eqs.~(\ref{eq:EQ017}). Only $\Psi_{\rm \RNum{1}}(y)$ is used in Kinnison's book (as well as in numerous other books and articles); this `convenient' choice 
bypasses the evaluation of the parameter $\zeta$ which enters the two other forms.
\item The ordered data $x_i$ (vertical axis) are plotted against the $k_i$ values (horizontal axis).
\item Ordinary linear regression on the ($k_i,x_i$) datapoints (for details, see the appendices in Ref.~\cite{Matsinos2023a}) yields estimates for the two parameters $a$ and $b$ of the linear relation: $x=a k + b$.
\item The parameters $a$ and $b$ are identified with the parameters $\alpha$ and $\mu$ of Eq.~(\ref{eq:EQ017_5}), respectively.
\end{itemize}
As Kinnison comments, a departure of the probability plot from the straight line is suggestive either of a problematical data acquisition or of the wrong class being used in the description of the extreme 
values. One remark is due: however useful they might be as guiding tools, empirical techniques frequently give results which differ - sometimes sizeably - from those obtained via the application of rigorous 
statistical procedures (e.g., from those resting upon an optimisation scheme).

The second option involves the extraction of estimates for the parameters $\mu$ and $\alpha$ of Eq.~(\ref{eq:EQ017_5}) from the sample mean $\avg{x}$ and the unbiased variance $s^2$ of the values $x_i$ of the 
reference dataset.
\begin{itemize}
\item From Eq.~(\ref{eq:EQ017_5}), one obtains $s^2 = \alpha^2 \, {\rm Var} (y)$. For the Gumbel class, the use of Eq.~(\ref{eq:EQ025}) leads to the relation $\alpha = \sqrt{6} s / \pi$.
\item Equation (\ref{eq:EQ017_5}) suggests that $\mu = \avg{x} - \alpha \avg{y}$. For the Gumbel class, the use of Eq.~(\ref{eq:EQ024}) leads to the relation $\mu = \avg{x} - \alpha \gamma$.
\end{itemize}

Although I might employ either of the two aforementioned options to obtain crude estimates for the parameters $\mu$ and $\alpha$ of Eq.~(\ref{eq:EQ017_5}), I mostly use the method of the Maximum-Likelihood 
Estimation (MLE) for reliable results~\footnote{The trouble with the application of the MLE method to optimisation problems is the lack of an established statistical procedure to assess the goodness-of-fit. 
One may generally compare the effectiveness of two PDFs in describing a set of input data (by resorting to the likelihood-ratio test), but there is no procedure for assessing the quality of \emph{one} fit 
(i.e., of one which assumes a specific underlying PDF): to summarise in one short sentence, there seems to be no way of knowing how `good' or how `bad' the `optimal' is.}. The PDF $\chi (x_i)$ of each 
independent and identically-distributed stochastic variable $x_i$ can be obtained from the corresponding PDF $\psi(y_i)$ of Section \ref{sec:MEVProperties}, i.e., $\psi_{\rm \RNum{1}}(y_i)$, 
$\psi_{\rm \RNum{2}}(y_i)$, or $\psi_{\rm \RNum{3}}(y_i)$, as the case might be.
\begin{equation} \label{eq:EQ032}
\chi (x_i) dx = \psi (y_i) dy \Rightarrow \chi (x_i) = \frac{1}{\alpha} \psi (y_i)
\end{equation}
The likelihood function $\mathcal{L} (x_1, x_2, \dots, x_n)$ is defined as the product of all PDFs $\chi (x_i)$.
\begin{equation} \label{eq:EQ033}
\mathcal{L} (x_1, x_2, \dots, x_n) = \prod_{i=1}^{n} \chi (x_i)
\end{equation}
To maximise the likelihood function, one therefore needs to maximise
\begin{equation} \label{eq:EQ034}
\ln \mathcal{L} (x_1, x_2, \dots, x_n) = \sum_{i=1}^{n} \ln \chi (x_i) = - n \ln \alpha + \sum_{i=1}^{n} \ln \psi (y_i) \, \, \, ,
\end{equation}
where each $y_i$ is obtained (for specific values of the parameters $\mu$ and $\alpha$) from the corresponding extreme value $x_i$ via Eq.~(\ref{eq:EQ017_5}).

For the Gumbel class, Eq.~(\ref{eq:EQ034}) takes the form
\begin{equation} \label{eq:EQ035}
\ln \mathcal{L} (x_1, x_2, \dots, x_n) = - n \ln \alpha - \sum_{i=1}^{n} \left( y_i + e^{-y_i} \right) \, \, \, ,
\end{equation}
and the maximisation of the likelihood function is equivalent to the minimisation of the function
\begin{equation} \label{eq:EQ036}
{\rm MF} = n \ln \alpha + \sum_{i=1}^{n} \left( y_i + e^{-y_i} \right) \, \, \, .
\end{equation}

For the Fr{\'e}chet class, Eq.~(\ref{eq:EQ034}) takes the form
\begin{equation} \label{eq:EQ037}
\ln \mathcal{L} (x_1, x_2, \dots, x_n) = n \ln \zeta - n \ln \alpha - \sum_{i=1}^{n} \left( y_i^{-\zeta} + (\zeta + 1) \ln y_i \right) \, \, \, ,
\end{equation}
and the maximisation of the likelihood function is equivalent to the minimisation of the function
\begin{equation} \label{eq:EQ038}
{\rm MF} = n \ln \alpha - n \ln \zeta + \sum_{i=1}^{n} \left( y_i^{-\zeta} + (\zeta + 1) \ln y_i \right) \, \, \, .
\end{equation}

For the Weibull class, Eq.~(\ref{eq:EQ034}) takes the form
\begin{equation} \label{eq:EQ039}
\ln \mathcal{L} (x_1, x_2, \dots, x_n) = n \ln (-\zeta) - n \ln \alpha - \sum_{i=1}^{n} \left( (-y_i)^{-\zeta} + (\zeta + 1) \ln (-y_i) \right) \, \, \, ,
\end{equation}
and the maximisation of the likelihood function is equivalent to the minimisation of the function
\begin{equation} \label{eq:EQ040}
{\rm MF} = n \ln \alpha - n \ln (-\zeta) + \sum_{i=1}^{n} \left( (-y_i)^{-\zeta} + (\zeta + 1) \ln (-y_i) \right) \, \, \, .
\end{equation}

Last but not least, only the frequency distribution of the measurements $x$ is sometimes available, i.e., the number of events $n_i$ (also known as `number of occurrences', `absolute frequencies', or simply 
`counts') collected in the $i$-th bin of a histogram containing $k$ consecutive, non-overlapping bins (not necessarily of the same width). The endpoints of the $i$-th bin of such a histogram are: $h_{i-1}$ 
and $h_{i}$, $i \in \mathbb{N}^+_k$, and the bin contains the number of events for which the measurements $x$ satisfy~\footnote{A question of convention in this study: when a value $x$ coincides with an endpoint, 
it is assigned to the bin on its right.}: $x \in [h_{i-1},h_{i})$. For specific values of the parameters $\mu$ and $\alpha$, one first obtains the probability $p_i$ that a measurement be collected in the 
$i$-th histogram bin.
\begin{equation} \label{eq:EQ041}
p_i = P(h_{i-1} \leq x < h_{i}) = \Psi \left( \frac{h_i - \mu}{\alpha} \right) - \Psi \left( \frac{h_{i-1} - \mu}{\alpha} \right) \, \, \, ,
\end{equation}
where $\Psi$ stands for any of the three functions of Eqs.~(\ref{eq:EQ017}).

Regarding the optimisation, two options emerge, both requiring knowledge of the number of events $n_i$ (i.e., not only of the relative frequencies which are associated with the empirical PDF).
\begin{itemize}
\item One may maximise the overall likelihood
\begin{equation} \label{eq:EQ042}
\mathcal{L} \coloneqq \prod_{i=1}^{k} p_i^{n_i} \, \, \, ,
\end{equation}
which is mathematically equivalent to minimising
\begin{equation} \label{eq:EQ043}
- \ln \mathcal{L} = - \sum_{i=1}^{k} n_i \ln p_i \, \, \, .
\end{equation}
\item One may minimise a conventional $\chi^2$ function, defined in this problem as
\begin{equation} \label{eq:EQ044}
\chi^2 = \sum_{i=1}^k \frac{ \left( n_i - N p_i \right)^2}{N p_i (1-p_i)} \, \, \, ,
\end{equation}
where the total number of occurrences $N$ is obviously given by
\begin{equation} \label{eq:EQ045}
N = \sum_{i=1}^k n_i \, \, \, ,
\end{equation}
\end{itemize}
The denominator of the ratio on the right-hand side of Eq.~(\ref{eq:EQ044}) is nothing but the variance of the binomial distribution.

\section{\label{sec:Applications}Applications}

To my knowledge, the EVT was first applied to a real-world problem by Gumbel in 1941 \cite{Gumbel1941}. In that study, the author set out to study the floods caused by the Rh{\^o}ne and Mississippi rivers: 
in the former case, predictions were extracted from the historical measurements, spanning over one century. In Table \RNum{3} of his paper, Gumbel lists the measurements of the discharges (i.e., of the 
volumetric flow rates) $x_i$ of the Rh{\^o}ne river in the Lyon area~\footnote{The data, $111$ entries, refer to the largest daily discharges in each year between 1826 and 1936.}, which were used for 
extracting estimates for the return period of floods, also including two sums in the paper ($\sum_i x_i=276\,773$ and $\sum_i x_i^2=744\,538\,565$, where the $x_i$'s are expressed in m$^3$ s$^{-1}$). 
However, using the values of Table \RNum{3} of Gumbel's paper, one obtains slightly different results ($\sum_i x_i=276\,762$ and $\sum_i x_i^2=744\,007\,294$). It is uncertain at this time whether the 
discrepancy is due to erroneous values in the table or to a miscalculation of both sums quoted in Gumbel's paper. I had hoped that I could clarify this question after gaining access to the original data, 
which had appeared in a 1938 paper by Aim{\'e} Coutagne (1882-1970) \cite{Coutagne1938}, a French Hydrology engineer. To this end, I requested Coutagne's paper from the only (or so it seems) library in 
France ({\'E}cole Sp{\'e}ciale des Travaux Publics, Cachan), where his report appears to be available, intending to correct the values (if indeed corrections are called for) and make them publicly available; 
however, it seems that this could not be possible without my physical presence in Cachan. As a result, the assumption in Section \ref{sec:Floods} is that the discrepancy in Gumbel's paper originates from a 
miscalculation of both sums quoted in his paper. In any case, even if the discrepancies are due to typographical mistakes in the contents of Table \RNum{3} of the 1941 paper, it is unlikely that the 
important results are affected beyond the quoted uncertainties.

Before indiscriminately applying the mathematical formalism to datasets, a word of caution is due. The development of the formalism in Section \ref{sec:MEVGeneralFormalism} rested upon the employment of the 
three limiting forms of Eqs.~(\ref{eq:EQ017}). However, it must be understood that nowhere does the formalism explicitly take the sample size $n$ into account. Therefore, the working hypothesis is that the 
sample size $n$ is `large enough' that the results of the analysis do not sizeably depend on it; unfortunately, it is not clear what `large enough' implies. To my knowledge, the subject of the `finite-$n$ 
effects', which could be better studied using simulated data, has not been addressed beyond the sketchy and inadequate level of Chapter 10 of Kinnison's book, see also Appendix \ref{App:AppB}. This is not an 
issue of minor importance, in particular when considering the \emph{very} slow convergence - towards $\Psi_{\rm \RNum{1}}(y)$ - of the CDF of the largest value of normally-distributed data, a critical remark 
first made by Fisher and Tippett \cite{Fisher1928} and understood after considering the expansion of Eq.~(\ref{eq:EQ020}).

As a result of the frequent dependence of the distributions (from which the samples are drawn) on time, it is generally difficult to find suitable applications of the EVT in Physics; the independence of the 
measurements is one additional, albeit lesser, problem. Regarding the former issue, the measurements are affected in two ways: first, bias may be introduced into the analyses because of the temporal evolution 
of the physical system under observation; second, the bias may pertain to developments in the instrumentation or to modifications in the measurement techniques over time. There is little doubt that most 
historical measurements suffer from at least one of these effects~\footnote{For instance, as a result of the climate crisis, the statistical analysis of the measurements, acquired from the observation of 
extreme weather phenomena (extreme rainfall, extreme drought, extreme temperature), is not straightforward. The main difficulty relates to the time-dependent definition of the word `extreme': events which 
were considered extreme in the first half of the $20$-th century are currently viewed as ordinary. In essence, the first of the two assumptions, which have been mentioned at the beginning of Section 
\ref{sec:DistributionNumberOfExceedances} (namely that all measurements are assumed drawn from the same distribution), is - as a result of the rising global average temperature over the course of the past 
five decades - not fulfilled.}. To study many physical phenomena, without regard to the time dependence of the corresponding distributions, would be tantamount to comparing financial data, acquired over 
several decades, without adjusting the values for inflation.

For the numerical minimisation, the MINUIT software package \cite{MINUIT} of the CERN library (FORTRAN version) has been used (in FORTRAN-callable mode), see also Appendices B and C of Ref.~\cite{Matsinos2023a}. 
As made clear in Ref.~\cite{MINUIT}, p.~4, the MINOS $1 \sigma$ uncertainties for optimisations based on the MLE method must involve the distance of $0.5$ to the extremum (SET ERR $0.5$). This is understood 
as follows. The PDF of a normally-distributed random variable $x$ reads as
\begin{equation} \label{eq:EQ045_01}
N(x; m, \sigma^2) = \frac{1}{\sigma \sqrt{2 \pi}} \, \exp \left( - \frac{(x-m)^2}{2 \sigma^2} \right) \, \, \, ,
\end{equation}
where the parameters $m$ and $\sigma^2$ denote the average and the variance of the distribution. The value of this PDF at the distance of $1 \sigma$ away from the average $m$ is equal to 
\begin{equation} \label{eq:EQ045_02}
N(m \pm \sigma; m, \sigma^2) = \frac{1}{\sigma \sqrt{2 \pi}} \, e^{-1/2} = N(m; m, \sigma^2) \, e^{-1/2} \, \, \, ,
\end{equation}
where $N(m; m, \sigma^2) = \left( \sigma \sqrt{2 \pi} \right) ^{-1}$ is evidently the maximal value of the normal PDF. Provided that the likelihood function $\mathcal{L}$ is approximately normal, the $1 \sigma$ 
uncertainties of the fitted values of the model parameters may be defined as corresponding to the level $\mathcal{L} = \mathcal{L}_{\rm max} e^{-1/2}$, which implies that $\ln \mathcal{L} = \ln \mathcal{L}_{\rm max} - 1/2$. 
As the optimisation involves the minimisation of the ${\rm MF} = - \ln \mathcal{L}$ of Eqs.~(\ref{eq:EQ036},\ref{eq:EQ038},\ref{eq:EQ040}), the uncertainties are obtained by increasing the minimal value 
${\rm MF_{min}}$ by $-(-1/2)=+1/2$. Using the definition of the conventional $\chi^2$ function, one can show that the aforementioned uncertainties involve the increase of $\chi^2_{\rm min}$ by $1$.

The routines CORSET and CORGEN of the CERN software library \cite{CERNLIB} (FORTRAN version) have been used in all Monte-Carlo simulations. In all cases, the PDFs of the physical quantities in question were 
first obtained (at the specific values of the independent variable); a few events in Section \ref{sec:Meteorites} had to be removed, as the simulation generated values outside the physical domain (negative 
crater diameters). The CDFs were subsequently obtained; the medians of the distributions were extracted via a cubic interpolation (and solution of the resulting cubic equations) around the CDF value of $0.5$. 
The reported CIs correspond to $1 \sigma$ effects in the normal distribution (${\rm CL} \approx 68.27$~\%): the lower endpoint of each CI corresponds to ${\rm CDF} \approx 0.158655$, the higher one to 
${\rm CDF} \approx 0.841345$. This procedure accounts for the possibility of asymmetrical (skewed) PDFs.

\subsection{\label{sec:Floods}On floods caused by the Rh{\^o}ne river}

\begin{figure}
\begin{center}
\includegraphics [width=15.5cm] {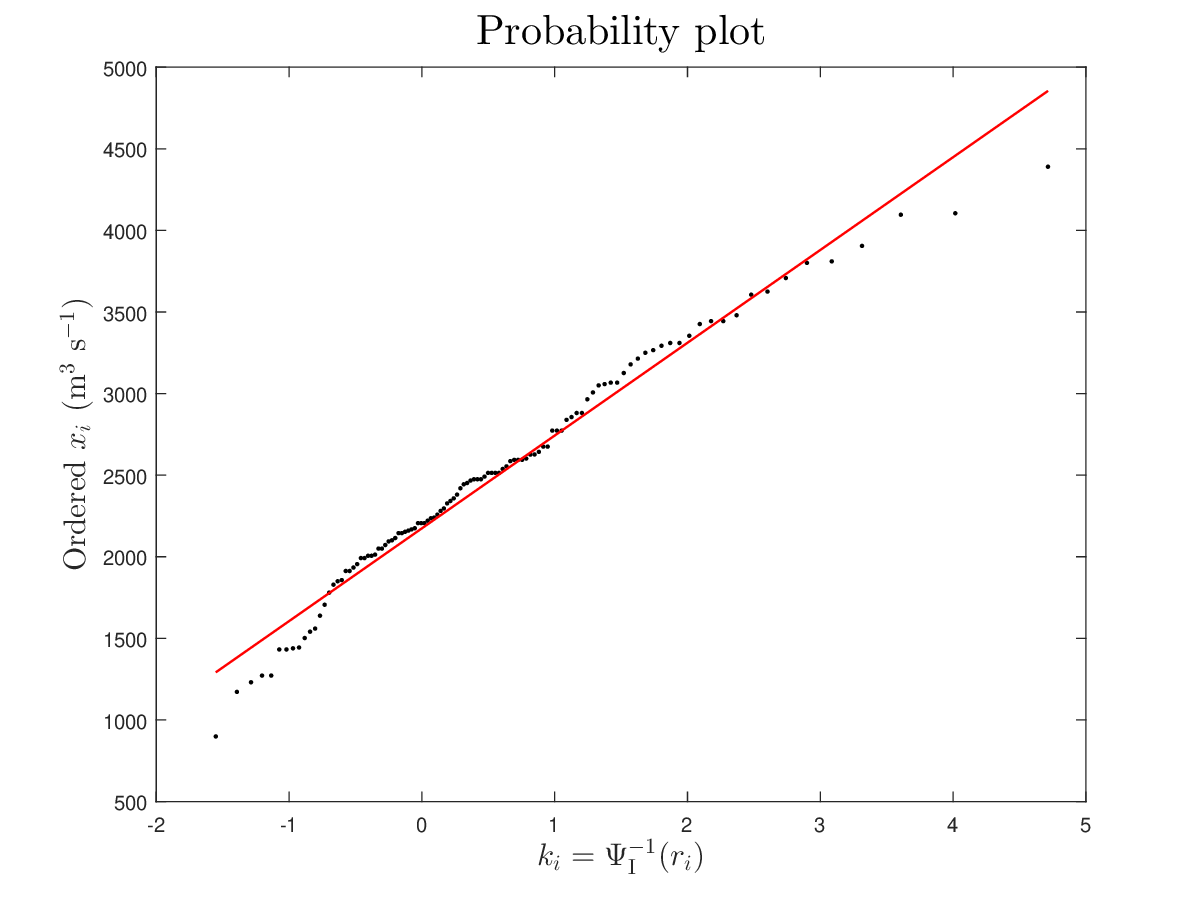}
\caption{\label{fig:ProbabilityPlotFloods}The probability plot for the historical measurements of the largest daily discharges $x_i$ of the Rh{\^o}ne river in each year between 1826 and 1936 \cite{Coutagne1938}. 
The values of the parameters $\mu$ and $\alpha$, obtained via ordinary linear regression, are equal to $2\,174$ m$^3$ s$^{-1}$ and $569$ m$^3$ s$^{-1}$, respectively. The optimal result of the linear regression 
is the red solid straight line.}
\vspace{0.5cm}
\end{center}
\end{figure}

The ordered set of the largest daily discharges $x_i$ of the Rh{\^o}ne river in each year between 1826 and 1936 \cite{Coutagne1938} are plotted against the uniform order statistic medians 
$k_i = \Psi^{-1}_{\rm \RNum{1}} (r_i)$ in Fig.~\ref{fig:ProbabilityPlotFloods}; the plotting positions $r_i$ have been obtained from Eq.~(\ref{eq:EQ031_5}) with $A=0$. The PCC between the arrays $x_i$ and 
$k_i$ comes out equal to about $0.9884$; the choice $A=3/8$ results in an imperceptible change: ${\rm PCC} \approx 0.9855$.

The minimisation of the function MF of Eq.~(\ref{eq:EQ036}) leads to the determination of the fitted parameter values and uncertainties of Table \ref{tab:ParametersFloods}. The off-diagonal element of the 
Hessian matrix is equal to about $0.327862$.

\vspace{0.5cm}
\begin{table}[h!]
{\bf \caption{\label{tab:ParametersFloods}}}The results of the optimisation to the measurements of the largest daily discharges $x_i$ of the Rh{\^o}ne river in each year between 1826 and 1936 
\cite{Coutagne1938}, as presented in Table \RNum{3} of Ref.~\cite{Gumbel1941}. The t-multiplier, corresponding to $1 \sigma$ effects in the normal distribution, has been applied to the quoted uncertainties. 
Gumbel's estimates (not extracted from an optimisation scheme, but from the sample mean and the unbiased variance of the input values $x_i$, see second option in Section \ref{sec:MEVAnalysis}) are given in 
Ref.~\cite{Coutagne1938} as $\mu = 2\,177.0$ m$^3$ s$^{-1}$ and $\alpha = 548.2$ m$^3$ s$^{-1}$, see Table \RNum{4}.
\vspace{0.25cm}
\begin{center}
\begin{tabular}{|l|c|c|}
\hline
Parameter & Fitted value & Corrected fitted\\
 & & uncertainty\\
\hline
\hline
$\mu$ (m$^3$ s$^{-1}$) & $2\,155$ & $64$\\
$\alpha$ (m$^3$ s$^{-1}$) & $636$ & $45$\\
\hline
\end{tabular}
\end{center}
\vspace{0.5cm}
\end{table}

\begin{figure}
\begin{center}
\includegraphics [width=15.5cm] {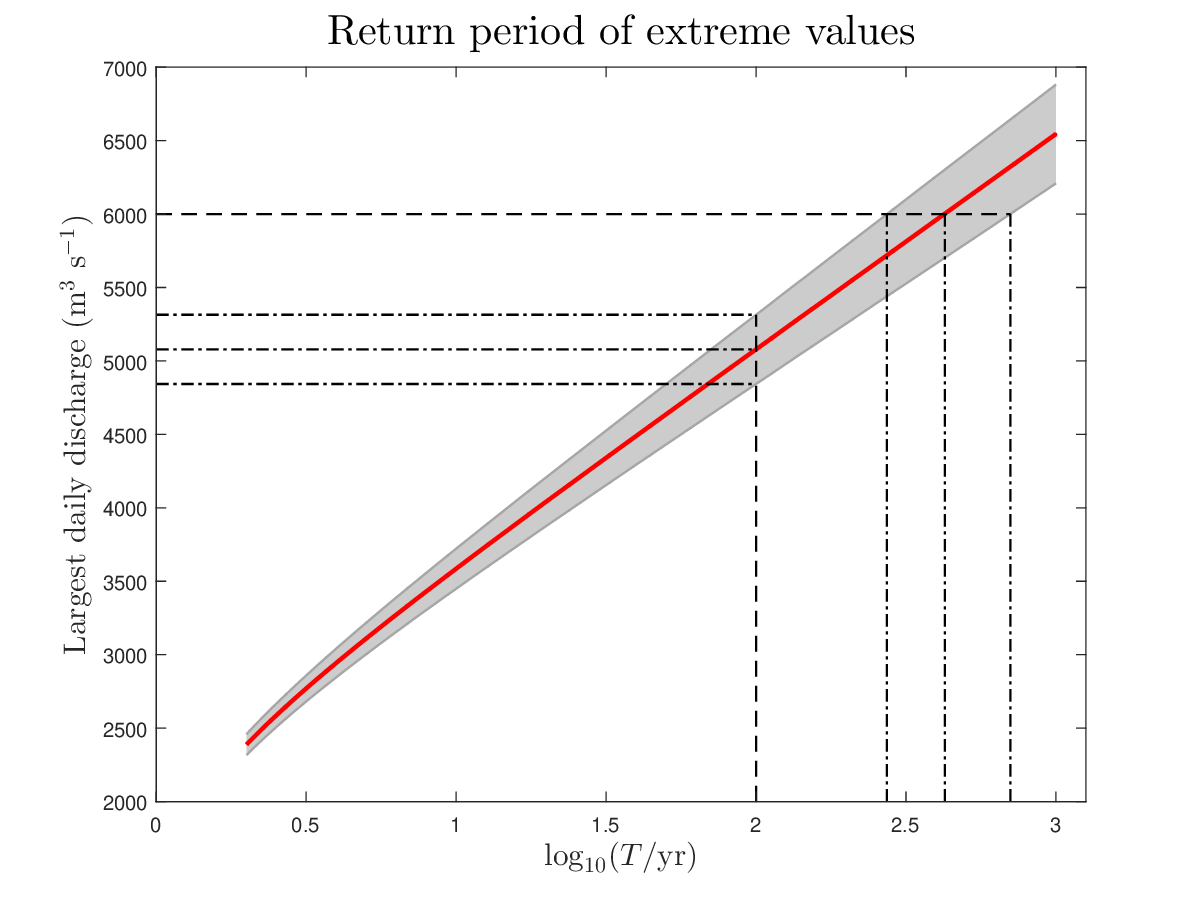}
\caption{\label{fig:ReturnPeriodFloods}The relation between the future (which, for the data analysed in this section, started on 1 January 1937) expectation values of the extreme discharges of the Rh{\^o}ne 
river and the (decimal logarithm of the) return period. The red solid curve represents median values, whereas the two grey solid curves delimit the CIs which are associated with $1 \sigma$ effects in the normal 
distribution (i.e., with ${\rm CL} \approx 68.27$~\%). To answer type-(a) questions (e.g., corresponding to a return period of $100$ yr, to be identified with the year 2036), one needs to determine the values 
of the ordinate corresponding to the intersection of the black dashed vertical straight line (e.g., at $\log_{10} (T/{\rm yr}) = 2$, $T$ in yr) and the red solid curve (for median values) or the two grey solid 
curves (for the determination of the $1 \sigma$ CIs). Similarly, to answer type-(b) questions (e.g., corresponding to a discharge of $6\,000$ m$^3$ s$^{-1}$), one needs to determine the values of the abscissa 
corresponding to the intersection of the black dashed horizontal straight line (e.g., at $6\,000$ m$^3$ s$^{-1}$) and the aforementioned curves.}
\vspace{0.5cm}
\end{center}
\end{figure}

Regarding the important predictions, resting upon the results of the optimisations in this study, one may pose two related questions.
\begin{itemize}
\item[a)] What severity of extraordinary events may be expected within a specific temporal (future) interval?
\item[b)] When are extraordinary events of a specific severity most likely to start occurring regularly?
\end{itemize}
These questions can be answered using the results of the same simulation, see Fig.~\ref{fig:ReturnPeriodFloods}. The expectation values of the extreme discharges of the Rh{\^o}ne river for five temporal 
(future, which - for the data analysed in this section - starts with the year 1937) intervals, given in Table \ref{tab:ReturnPeriodFloods}, have been obtained from $1$ million Monte-Carlo events per case 
using the results of Table \ref{tab:ParametersFloods}, along with the Hessian matrix of the MLE fit; of course, the quantity $\Delta t$ is equal to $1$ yr in this section. It should be mentioned that the 
predictions, which are based on the extreme-value analysis of the historical measurements, cannot possibly take later (i.e., introduced into the physical system under observation after the finalisation of 
the reference dataset) effects into account.

\vspace{0.5cm}
\begin{table}[h!]
{\bf \caption{\label{tab:ReturnPeriodFloods}}}The median expectation values of the extreme discharges of the Rh{\^o}ne river and the corresponding CIs which are associated with $1 \sigma$ effects in the normal 
distribution (i.e., with ${\rm CL} \approx 68.27$~\%) for five values of the return period $T$. The CDF $\Psi_{\rm \RNum{1}}(y)$ and the return period $T$ are related via Eq.~(\ref{eq:EQ017_4}). These results 
suggest that the median expectation value of the extreme discharges of the Rh{\^o}ne river will exceed the largest value of the historical measurements by $50$~\% by the year 2936. These predictions have been 
obtained from $1$ million Monte-Carlo events per case using the results of Table \ref{tab:ParametersFloods}, along with the Hessian matrix of the MLE fit.
\vspace{0.25cm}
\begin{center}
\begin{tabular}{|l|c|c|c|}
\hline
$\Psi_{\rm \RNum{1}}(y)$ & T (yr) & Median discharge (m$^3$ s$^{-1}$) & $\approx 68.27$~\% CI (m$^3$ s$^{-1}$)\\
\hline
\hline
$0.5$ & $2$ & $2\,388$ & $[2\,317,2\,459]$\\
$0.9$ & $10$ & $3\,585$ & $[3\,449,3\,722]$\\
$0.95$ & $20$ & $4\,043$ & $[3\,877,4\,209]$\\
$0.99$ & $100$ & $5\,079$ & $[4\,843,5\,314]$\\
$0.999$ & $1\,000$ & $6\,545$ & $[6\,209,6\,882]$\\
\hline
\end{tabular}
\end{center}
\vspace{0.5cm}
\end{table}

Flooding forecast is one of the domains of application of the EVT wherein, provided that the projected risk is appreciable, human intervention is bound to invalidate any predictions obtained from the 
extreme-value analysis: this intervention has several forms of long-term flood-prevention measures, from the improvement of the drainage infrastructure to the diversion of part of the water mass towards 
newly built waterways and/or storage areas. Once more, any predictions, obtained from the statistical analyses of measurements acquired between 1826 and 1936, can take no account of effects which became 
significant afterwards, e.g., as are those induced by global warming \cite{Ruiz2015}.

\subsection{\label{sec:Meteorites}On the meteorite impacts on the Earth}

Meteorites have impacted the surface of the Earth since our planet took shape, a few Gyr ago. The best-known event (and to this time third in rank in terms of severity), which gave rise to the Chicxulub 
crater, occurred about $66$ Myr ago, when a sizeable meteorite ($10$ to $15$ km in diameter) ended its wandering in the Cosmos by striking the Yucat{\'a}n Peninsula in Mexico. That event triggered 
earthquakes, tsunamis, and extensive wildfires, which (probably in parallel with concurrent severe volcanic activity and the release of poisonous gases, e.g., of SO$_2$) wiped out three-quarters of all 
plant and animal species on the Earth (Cretaceous–Paleogene extinction event).

Information about the events involving meteorite impacts on the Earth can be obtained from several sources, some of which pay little attention to `details', e.g., to the evidence for `shock metamorphism', a 
condition which necessitates the emergence of ultra-high pressures (typically, $\gtrapprox 10-20$ GPa) and temperatures (typically, $\gtrapprox 2000^\circ$C), as are those characterising the hyper-velocity 
collisions of meteorites with the Earth, see Table 4.1 of Ref.~\cite{French1998}. In contrast, endogenous metamorphism, involving terrestrial processes such as earthquakes, is associated with significantly 
lower pressures (generally, $\lessapprox 1-3$ GPa) and temperatures (generally, $\lessapprox 1000^\circ$C), e.g., see also Fig.~2 of Ref.~\cite{Stoffler2018}. It is my opinion that, regarding scientific 
purposes, few DBs of meteorite impacts contain reliable information: the data seem to enter the majority of the DBs without being adequately scrutinised, i.e., without evidence of high shock pressure on the 
material in the vicinity of the impact craters and structures, without regard for the physico-chemical properties of the objects in the vicinity of such structures, regardless of whether or not fragments of 
the purported meteorite have been recovered, and so on. One glaring oversight is that, despite the fact that the R{\'i}o Cuarto craters in Argentina - which were associated with meteorite impacts before 
2002 - are currently presumed to have originated from aeolian processes \cite{Cione2002}, they are still included in several lists of impact craters.

Although I did not manage to establish contact with the administrator and the research group responsible for the `Impact Earth Database' (IEDB) \cite{IEDB}, I decided to opt for using their data. At this 
time, details of $208$ impact craters are found in the IEDB: $20$ in Africa, $65$ in North America (including $1$ in Greenland, which geographically belongs to North America), $13$ in South America, $35$ 
in Asia (including the entirety of Russia), $32$ in Australia, and $43$ in Europe. One datapoint from the Australian set of structures (corresponding to the small Dalgaranga impact crater) was excluded on 
account of the incompatibility of the two estimates for its age; on this issue, the IEDB Group comment: ``A rough estimate of $<0.003$ Myr was provided by the preservation of crater morphology and 
stratigraphic age constraints (Shoemaker and Shoemaker, 1988). Other work, using the $^{10}$Be-$^{26}$Al exposure age, estimated $0.27$ Myr (Shoemaker \etal, 1990). The age is very uncertain.''

The IEDB is divided into three parts: a) a set of $194$ hyper-velocity impact craters, b) a set of $12$ impact craters, and c) a set of $43$ sites with impact deposits. The reason that the sum of their impact 
craters (hyper-velocity or otherwise) is $206$, rather than $208$, is because the IEDB Group do not include two of the craters in Russia (Macha and Sobolev), which \emph{are} present in their DB, in their 
lists of the aforementioned categories (a) and (b). A cursory examination of the IEDB reveals that four of the entries in the main block of the data (range of identifiers of the craters between $1$ and $206$) 
have been deleted: although it is unclear which these craters had been, they were associated with the identifiers $57$, $65$, $123$, and $189$. Perhaps of relevance to this observation is the omission from 
the IEDB of four impact craters (three fairly large and one small) which are generally considered `confirmed'. These impact craters are:
\begin{itemize}
\item the Gusev impact crater in Russia (with a rim diameter of $3$ km and an age of $49$ Myr) \cite{Gusev};
\item the Crawford impact crater in Australia (with a rim diameter of $8.5$ km and an age of $35$ Myr) \cite{Crawford};
\item the Piccaninny impact crater in Australia (with a rim diameter of $7$ km and an age of $180$ Myr) \cite{Piccaninny}; and
\item the Ilumetsa impact crater in Estonia (with a rim diameter of $80$ m and an age of about $7.1$ kyr) \cite{Ilumetsa}.
\end{itemize}
My efforts notwithstanding, a veil of mystery covers the six aforementioned structures: in spite of my two attempts at establishing contact with the IEDB Group (on 29 and 31 May 2024), I received no feedback 
from them. Unwilling to discard data without a reason, I decided to enhance my initial DB of $207$ datapoints, by including the data corresponding to the four aforementioned craters after fixing their 
properties (i.e., the diameter of each structure and the temporal separation to the time instant of the corresponding impact) from other sources. I stress that, as a result of the procedure in the analysis - 
i.e., of selecting the largest value in each histogram bin (see below), it matters not whether one includes the six aforementioned entries in the analysis DB or not: their inclusion is nevertheless a matter 
of principle.

I shall next outline the procedure employed in the data analysis.
\begin{itemize}
\item Each crater $i$ of the analysis DB is represented by one datapoint in a two-dimensional space: ($\tilde{d}_i$,$t_i$).
\begin{itemize}
\item The apparent crater diameter $\tilde{d}_i$ (expressed in km) ``represents the diameter of the outermost ring of (semi-)continuous concentric normal faults, measured with respect to the pre-impact surface 
(i.e., accounting for the amount of erosion that has occurred). For the majority of impact structures on Earth this will be the only measurable diameter.'' \cite{CraterDiameter}
\item The quantity $t_i$ (expressed in Myr) represents an estimate for the age of the $i$-th structure, i.e., for the temporal interval between the time instant of the meteorite impact and the current time.
\end{itemize}
\item The crater diameters $\tilde{d}_i$ are histogrammed in bins of the age. The optimal choice of the \emph{constant} bin width is a trade-off between a fine resolution (accompanied by the largest length 
of the contiguous bins containing non-zero counts) and information loss (ensuing from the use of bins which are unnecessarily wide). In this problem, the choice of a bin width of $20$ Myr enables the analysis 
of the data up to $320$ Myr (i.e., from $320$ Myr ago to the current time), see Table \ref{tab:HistogramMeteorites}; there is no structure with age $t_i$ between $320$ and $340$ Myr.
\item Histogramming the data essentially transforms the original array $\tilde{d}_i$ into a $2 \times 2$ matrix $\tilde{d}_{ij}$, assigning to each of the values one histogram bin ($i$) and one identifier 
($j$) within that bin. The $i$-th histogram bin contains $N_i$ values.
\item The largest of the $\tilde{d}_{ij}$ values for $j \in \mathbb{N}^+_{N_i}$ is selected in each histogram bin $i \in \mathbb{N}^+_{16}$, as the histogram of Table \ref{tab:HistogramMeteorites} comprises 
$16$ bins:
\begin{equation} \label{eq:EQ046}
d_i = \max_{j \in \mathbb{N}^+_{N_i}} \tilde{d}_{ij} \, \, \, .
\end{equation}
\end{itemize}

\vspace{0.5cm}
\begin{table}[h!]
{\bf \caption{\label{tab:HistogramMeteorites}}}The histogram of the apparent crater diameter in bins of the age. The quantity $N_i$ represents the number of counts (meteorite impacts) in the $i$-th histogram 
bin. The largest apparent crater diameter is selected in each histogram bin, see Eq.~(\ref{eq:EQ046}), and - to enable the analysis of the data and the extraction of predictions - it is assigned to the centre 
of that bin. The histogram contains $145$ of the original $211$ datapoints ($\tilde{d}_i$,$t_i$) of the analysis DB; the ages of the remaining structures exceed $340$ Myr.
\vspace{0.25cm}
\begin{center}
\begin{tabular}{|c|c|c|}
\hline
Bin endpoints (Myr) & Number of & $d_i$ (km)\\
 & counts $N_i$ &\\
\hline
\hline
$0.0-20.0$ & $48$ & $24.0$\\
$20.0-40.0$ & $13$ & $100.0$\\
$40.0-60.0$ & $11$ & $45.0$\\
$60.0-80.0$ & $8$ & $180.0$\\
$80.0-100.0$ & $9$ & $13.0$\\
$100.0-120.0$ & $3$ & $13.5$\\
$120.0-140.0$ & $6$ & $84.0$\\
$140.0-160.0$ & $10$ & $70.0$\\
$160.0-180.0$ & $5$ & $19.0$\\
$180.0-200.0$ & $8$ & $80.0$\\
$200.0-220.0$ & $5$ & $100.0$\\
$220.0-240.0$ & $3$ & $40.0$\\
$240.0-260.0$ & $5$ & $40.0$\\
$260.0-280.0$ & $2$ & $32.0$\\
$280.0-300.0$ & $7$ & $36.0$\\
$300.0-320.0$ & $2$ & $4.5$\\
\hline
\end{tabular}
\end{center}
\vspace{0.5cm}
\end{table}

A word of caution is due. One of the problems when analysing data acquired via the observation of physical phenomena bears on the randomness of occurrence of such events: evidently, earthquakes and meteorite 
impacts care not for regularity of occurrence. The flood data, analysed in Section \ref{sec:Floods}, also suffer from randomness, though (given the seasonality of the weather patterns) there is more 
regularity in the time of occurrence of specific weather phenomena in each calendar year: for instance, Octobers in Geneva, Switzerland, bring the largest amount of precipitation and Januaries the largest 
amount of snow. To be able to analyse the physical data and come up with predictions, one needs to assume the regularity of occurrence of the extraordinary events, in practice assigning the selected value 
(e.g., the $d_i$'s of Table \ref{tab:HistogramMeteorites}) to the centre of the corresponding histogram bins.

Before continuing, another remark is due. The severity of an impact will be judged by the apparent rim diameter of the impact crater, the only reliable observable. The size of the apparent rim diameter 
depends on numerous factors: predominantly, on the kinetic energy and on the angle of incidence of the projectile/impactor (i.e., of the meteorite) when it struck the ground, but also on the time of impact 
$t_i$, as well as on the topographical details of the impact site: for instance, erosion in its numerous forms (precipitation, running water, wind, formation of a lake inside the impact crater, etc.) and 
burial diagenesis, including the effects of volcanic activity in the vicinity of the impact crater, hinder the extraction of reliable results when studying the properties of the impact craters. Dence puts 
forward a rough relation between the final diameter $D$ (expressed in km) of an impact crater and the energy $E$ (expressed in J), released during the meteorite impact (estimated from the level of shock 
metamorphism at the centre of the structure), as: $D = a E^b$, where $a \approx 2.87 \cdot 10^{-5}$ and $b \approx 1/3.44$ \cite{Dence2006}. Using this relation, one could come up with estimates for the 
released energy $E$ in each impact, and analyse those values (rather than the crater diameters $d_i$ of Table \ref{tab:HistogramMeteorites}); this possibility has not been pursued in this study.

The next question involves the choice of the class of functions which best accounts for the distribution of the extreme values $d_i$ of Table \ref{tab:HistogramMeteorites}, see Eqs.~(\ref{eq:EQ017}). To 
provide an answer to this question, I shall propose (for the Fr{\'e}chet and Weibull classes, which - in comparison with the Gumbel class - have one additional DoF) the maximisation of the PCC between
\begin{itemize}
\item the uniform order statistic medians $k_i \coloneqq \Psi^{-1} (r_i)$, where the normalised-rank array $r_i$ has been defined in Eq.~(\ref{eq:EQ031_5}) and the three forms of the quantile function $\Psi^{-1}$ 
can easily be obtained from Eqs.~(\ref{eq:EQ017}), and
\item the measurements $d_i$ arranged in ascending order.
\end{itemize}
The PCC between the arrays $k_i$ and (the ordered values) $d_i$ is straightforward to obtain for the Gumbel class, whereas the $\zeta$-dependence of the $k_i$ values in the two other cases somewhat obfuscates 
the evaluation. To remove this additional DoF in case of the Fr{\'e}chet and the Weibull classes, $\zeta$ may be varied until the PCC is maximised. I am not aware of works in the literature where this 
procedure has been put forward or into application. In any case, the fixation of the parameter $\zeta$ from the probability plot is a welcoming prospect for one additional reason: it turns a three-parameter 
fit into a two-parameter fit, preventing `fit drifting', a situation hampering the extraction of reliable results in optimisation problems where the correlations among the model parameters turn out to be 
large (be that the result of the inherent properties of the mathematical modelling and/or of the scarcity of the input data).

\begin{figure}
\begin{center}
\includegraphics [width=15.5cm] {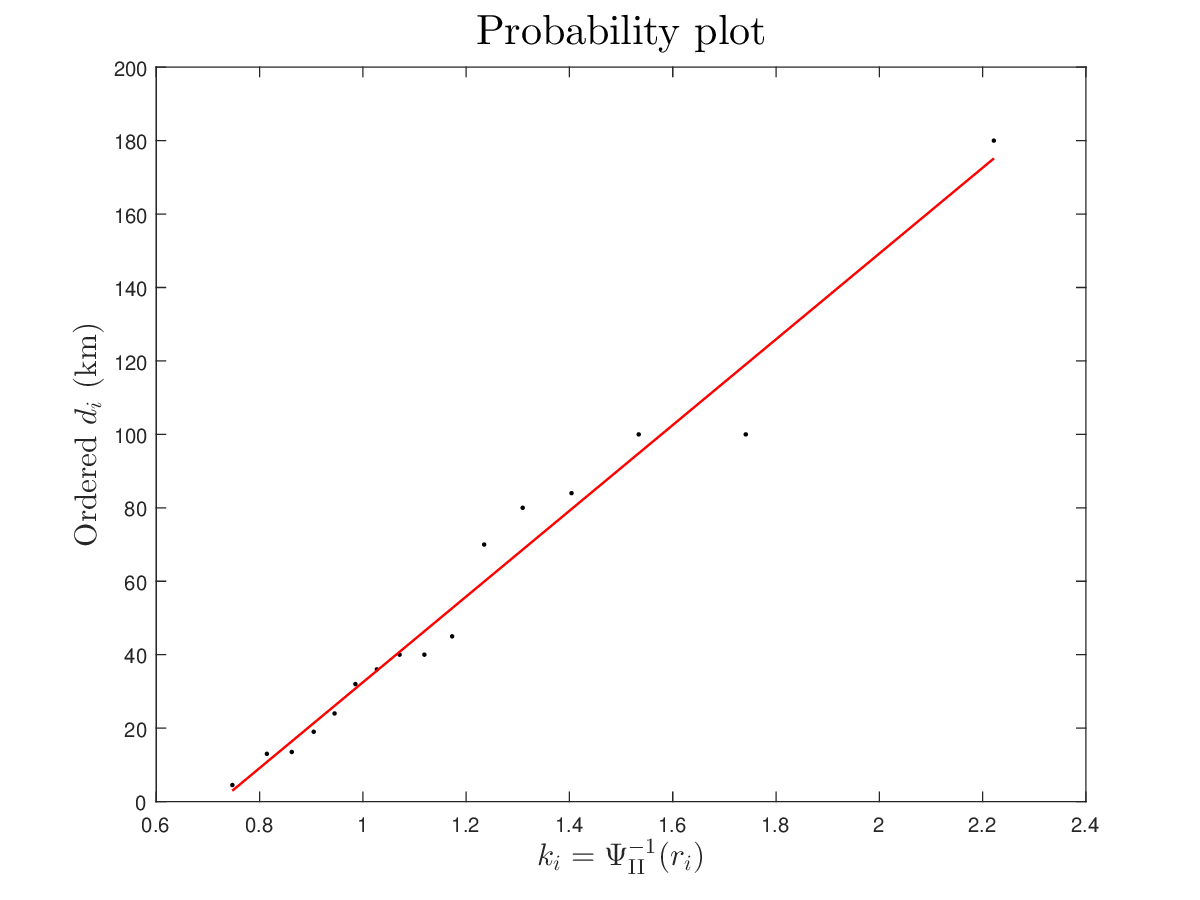}
\caption{\label{fig:ProbabilityPlotMeteorites}The probability plot for the impact diameters $d_i$ of Table \ref{tab:HistogramMeteorites}. The values of the parameters $\mu$ and $\alpha$, obtained via ordinary 
linear regression (the fitted results of which yielded the red solid straight line), are equal to $-84$ km and $117$ km, respectively.}
\vspace{0.5cm}
\end{center}
\end{figure}

The PCC values, corresponding to the use of the Gumbel and Fr{\'e}chet classes in the description of the crater diameters $d_i$, are given in Table \ref{tab:PCCMeteorites} for two choices of the quantity $A$ which 
enters the definition of the normalised-rank array $r_i$, see Eq.~(\ref{eq:EQ031_5}). The choice $A=3/8$ yields slightly larger PCC values for both classes. In addition, the employment of the Fr{\'e}chet 
class in the description of the data yields slightly larger PCC values. On the basis of the results of Table \ref{tab:PCCMeteorites}, the Fr{\'e}chet class will be used in the description of the extreme 
values of Table \ref{tab:HistogramMeteorites}, along with $A=3/8$ in the probability plot of Fig.~\ref{fig:ProbabilityPlotMeteorites}, and $\zeta=4.056917$ both in the probability plot and in the optimisation 
pertinent to this section.

\vspace{0.5cm}
\begin{table}[h!]
{\bf \caption{\label{tab:PCCMeteorites}}}The PCC values between the uniform order statistic medians $k_i$ and the ordered diameters $d_i$ of Table \ref{tab:HistogramMeteorites} for the Gumbel and 
Fr{\'e}chet classes, and two popular choices of the parameter $A$ associated with the normalised-rank array $r_i$ (hence also with $k_i$), see Eq.~(\ref{eq:EQ031_5}). For the Fr{\'e}chet class, the quoted 
result for the exponent $\zeta$ maximises the PCC; the Gumbel class does not contain any such parameter. The PCC (considered a function of $\zeta$), obtained for the Weibull class, is monotonic (increasing) 
as $\zeta$ decreases from $0^-$ to $-\infty$, asymptotically approaching the result quoted for the Gumbel class.
\vspace{0.25cm}
\begin{center}
\begin{tabular}{|c|c|c|}
\hline
Class & PCC & $\zeta$\\
\hline
\hline
\multicolumn{3}{|c|}{Results for $A=0$}\\
\hline
Gumbel & $0.9662$ & $-$\\
Fr{\'e}chet & $0.9857$ & $2.834029$\\
\hline
\multicolumn{3}{|c|}{Results for $A=3/8$}\\
\hline
Gumbel & $0.9734$ & $-$\\
Fr{\'e}chet & $0.9872$ & $4.056917$\\
\hline
\end{tabular}
\end{center}
\vspace{0.5cm}
\end{table}

The minimisation of the function MF of Eq.~(\ref{eq:EQ038}) led to the determination of the fitted parameter values and uncertainties of Table \ref{tab:ParametersMeteorites}. The off-diagonal element of the 
Hessian matrix is equal to about $-0.967858$.

\vspace{0.5cm}
\begin{table}[h!]
{\bf \caption{\label{tab:ParametersMeteorites}}}The results of the optimisation to the impact diameters $d_i$ of Table \ref{tab:HistogramMeteorites}. The t-multiplier, corresponding to $1 \sigma$ effects in 
the normal distribution, has been applied to the quoted uncertainties. The parameter $\zeta$ has been fixed from the maximisation of the PCC, see Table \ref{tab:PCCMeteorites}, result for $A=3/8$.
\vspace{0.25cm}
\begin{center}
\begin{tabular}{|l|c|c|}
\hline
Parameter & Fitted value & Corrected fitted\\
 & & uncertainty\\
\hline
\hline
$\mu$ (km) & $-73$ & $23$\\
$\alpha$ (km) & $105$ & $27$\\
\hline
\end{tabular}
\end{center}
\vspace{0.5cm}
\end{table}

\begin{figure}
\begin{center}
\includegraphics [width=15.5cm] {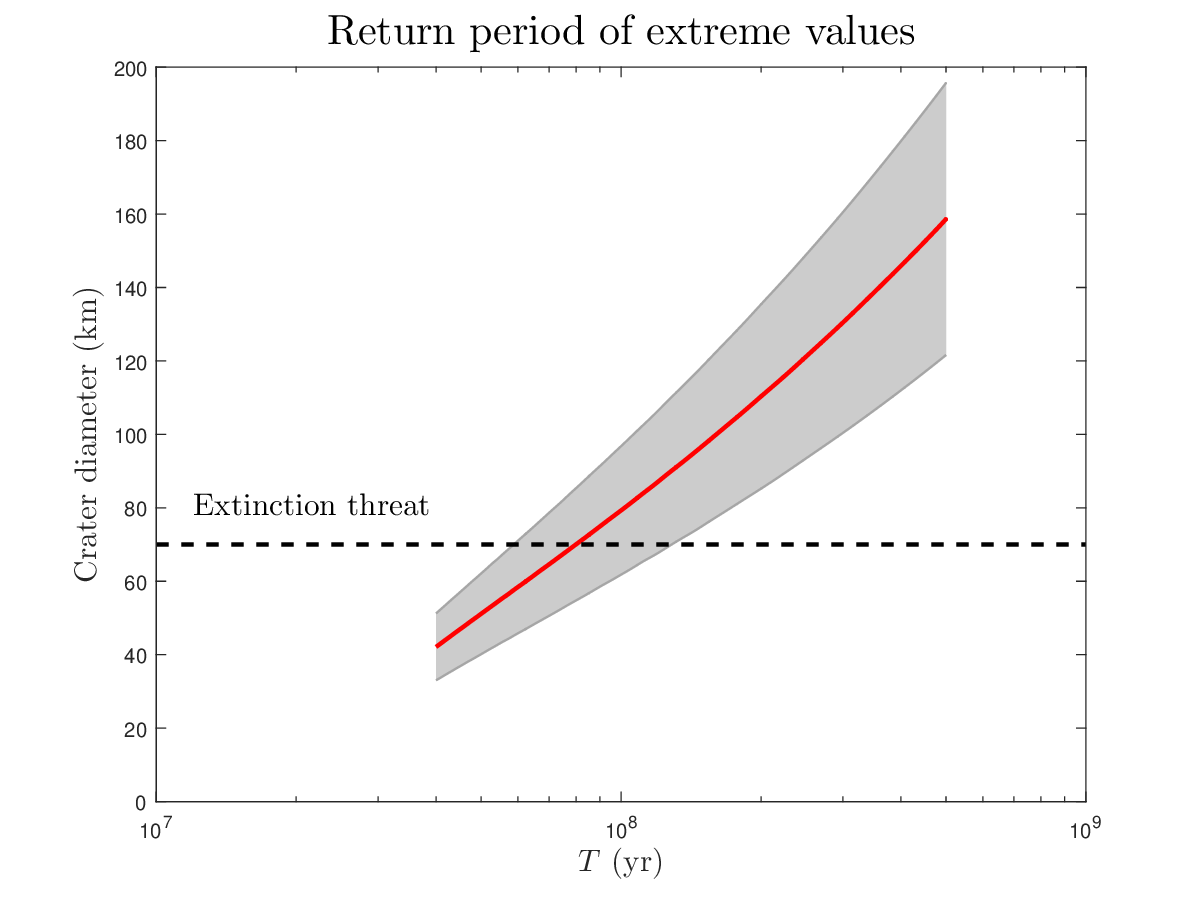}
\caption{\label{fig:ReturnPeriodMeteorites}The relation between the expectation values of the future extreme impact craters and the return period. The red solid curve represents median values, whereas the 
two grey solid curves delimit the CIs which are associated with $1 \sigma$ effects in the normal distribution (i.e., with ${\rm CL} \approx 68.27$~\%). The black dashed horizontal straight line is associated 
with a crater diameter of $70$ km, which is currently considered to represent the approximate outset of extinction threat, see also Table \ref{tab:ProbabilityOfOccurrenceMeteorites}.}
\vspace{0.5cm}
\end{center}
\end{figure}

The expectation values of the extreme impact diameters for three temporal (future) intervals, given in Table \ref{tab:ReturnPeriodMeteorites} and shown in Fig.~\ref{fig:ReturnPeriodMeteorites}, have been 
obtained from $1$ million simulated events per case using the results of Table \ref{tab:ParametersMeteorites} for the parameters $\mu$ and $\alpha$, along with the Hessian matrix of the MLE fit. The results of 
the simulation of this work, schematically shown in Fig.~\ref{fig:ReturnPeriodMeteorites}, do not support the frequently-voiced claim that Chicxulub-sized craters occur every $100$ Myr. It may be that the 
justification for this claim rests upon one sentence, which can be found in the important contribution by Grieve and Shoemaker to the book \emph{Hazards due to Comets and Asteroids}; on p.~452 of that book, 
the authors write: ``From estimates of the average terrestrial cratering rate, K/T-sized impact events occur on time scales of $\sim 100$ Myr.'' \cite{Grieve1991} Several authors consider this statement 
\emph{absolute}, whereas Grieve and Shoemaker meant it as \emph{approximate} (`order-of-magnitude-wise').

\vspace{0.5cm}
\begin{table}[h!]
{\bf \caption{\label{tab:ReturnPeriodMeteorites}}}The median expectation values of the extreme impact diameters $d_i$ and the corresponding CIs which are associated with $1 \sigma$ effects in the normal 
distribution (i.e., with ${\rm CL} \approx 68.27$~\%) for three values of the return period $T$, obtained from $1$ million simulated events per case using the results of Table \ref{tab:ParametersMeteorites}, 
along with the Hessian matrix of the MLE fit. The CDF $\Psi_{\rm \RNum{2}}(y)$ and the return period $T$ (expressed in Eq.~(\ref{eq:EQ017_4}) as a multiple of $\Delta t = 20$ Myr, see Table 
\ref{tab:HistogramMeteorites}) are related via Eq.~(\ref{eq:EQ017_4}).
\vspace{0.25cm}
\begin{center}
\begin{tabular}{|l|c|c|c|}
\hline
$\Psi_{\rm \RNum{2}}(y)$ & T (Myr) & Median crater & $\approx 68.27$~\% CI (km)\\
 & & diameter (km) & \\
\hline
\hline
$0.5$ & $40$ & $42$ & $[33,51]$\\
$0.9$ & $200$ & $110$ & $[85,135]$\\
$0.95$ & $400$ & $146$ & $[112,180]$\\
\hline
\end{tabular}
\end{center}
\vspace{0.5cm}
\end{table}

This work suggests that ``K/T-sized impact events'' are considerably less frequent than it is believed. There are only two craters in the analysis DB of this work, whose diameters exceed the one of the 
Chicxulub crater: Vredefort, South Africa ($300$ km) and Sudbury, Canada ($200$ km); the diameters of the fourth and fifth largest on the list (Manicouagan, Canada and Popigai, Russia) are nearly half ($100$ 
km) of the diameter of the Chicxulub crater. Therefore, if one means `events associated with an impact crater at least equal to the one which was generated after the K/T-sized impact event' when talking 
about ``K/T-sized impact events,'' then there are only three craters in total to consider, the largest three of all craters in the analysis DB. These three meteorite impacts (chronologically ordered) yield 
two temporal separations, which differ by one order of magnitude, one being equal to about $1.8$ Gyr (temporal interval between the formations of the - third largest - Chicxulub and - second largest - Sudbury 
craters), the other to about $173.5$ Myr (temporal interval between the formations of the - second largest - Sudbury and the - largest - Vredefort craters). Both temporal separations exceed the ``time scales 
of $\sim 100$ Myr;'' and, undoubtedly, the temporal separation between the third and second largest events is nowhere close to ``$\sim 100$ Myr.'' This is one of my arguments against the misconception that 
Chicxulub-sized craters occur (on average) every $100$ Myr; as Fig.~\ref{tab:ParametersMeteorites} suggests, such extraordinary events are significantly less frequent.

I believe that it makes no sense to extract predictions (in Table \ref{tab:ReturnPeriodMeteorites} and in Fig.~\ref{fig:ReturnPeriodMeteorites}) extending to millions (let alone billions) of years, at a 
time when the survival of humankind through the next millennium is questionable. A more realistic question thus emerges: what is the probability that a serious meteorite impact occur within the next $100$ 
years? Predictions will be obtained from the fitted values and uncertainties of Table \ref{tab:ParametersMeteorites} for the parameters $\mu$ and $\alpha$, along with the Hessian matrix of the MLE fit. 
Before that, however, a meaningful (involving the critical risk levels) list of crater diameters must be compiled.

It is generally believed that meteorite impacts which result in crater diameters of the order of $1$ km cause local disruption, whereas those giving rise to crater diameters of the order of $10$ km affect 
areas representative of medium-sized countries. It is also believed that meteorite impacts resulting in crater diameters of about $40$ km have severe worldwide consequences, whereas those giving rise to 
crater diameters of about $70$ km wreak havoc. These comparisons are helpful in comprehending the severity of the Chicxulub event about $66$ Gyr ago. Fortunately for humankind, the frequency of occurrence 
of meteorite impacts decreases with increasing crater diameter. To put everything in perspective, meteorite impacts occurring once in $100$ Myr are expected to occur with a probability of $10^{-8}$ per 
annum, or of $10^{-6}$ per century (which is of order of magnitude of the human life expectancy at birth). Table \ref{tab:ProbabilityOfOccurrenceMeteorites} contains the results of the simulation of this 
work for six meteorite impacts associated with the range of crater diameters between $1$ and $180$ km. In comparison, the American Cancer Society expects that $19$~\% of the men and $17$~\% of the women in 
the USA will die from cancer \cite{ACS}.

One might assume the thesis that samples of dimension $16$ are too small for the procedure set forward in Section \ref{sec:MEVAnalysis}. In Chapter 10 of his book \cite{Kinnison1983}, Kinnison suggests the 
application of Order Statistics for the extraction of predictions from small samples, without defining which sample should be understood as `small'. Be that as it may, the application of Kinnison's 
recommendation to the data of Table \ref{tab:HistogramMeteorites} of this work resulted in a median return period of $472$ Myr for craters at least equal to the Chicxulub crater ($\approx 68.27$~\% CI of the 
return period: $184$ to $1862$ Myr). Therefore, the estimate for the return period of ``K/T-sized impact events,'' obtained from the application of Order Statistics to the data, also suggests that such events 
are less frequent that they are believed to be. For the sake of comparison with the last row of Table \ref{tab:ProbabilityOfOccurrenceMeteorites}, the median probability of occurrence of such an event is 
about $0.212 \cdot 10^{-6}$ per century ($\approx 68.27$~\% CI of the probability of occurrence per century: $0.054 \cdot 10^{-6}$ to $0.543 \cdot 10^{-6}$).

\vspace{0.5cm}
\begin{table}[h!]
{\bf \caption{\label{tab:ProbabilityOfOccurrenceMeteorites}}}The probability of occurrence of meteorite impacts within the next $100$ years, along with the extracted CIs (representing $1 \sigma$ effects in 
the normal distribution), for six values of the resulting crater diameter. These estimates have been obtained from $10$ million simulated events per case using the results of Table \ref{tab:ParametersMeteorites}, 
along with the Hessian matrix of the MLE fit. When I mention the word `extinction' in this table, I do not imply `extinction' of life on the Earth. I rather refer to the availability of physicists to write 
papers about the EVT and its applications. Mother Nature will always find a way forwards, with or without humans.
\vspace{0.25cm}
\begin{center}
\begin{tabular}{|l|l|c|c|}
\hline
$d$ (km) & Comment & Probability of occurrence & $\approx 68.27$~\% CI\\
 & & per century $\times 10^6$ & $\times 10^6$ \\
\hline
\hline
$1$ & Local disruption & $4.92$ & $[4.75,4.99]$\\
$10$ & Country disruption & $4.63$ & $[4.31,4.86]$\\
$40$ & Worldwide disruption & $2.63$ & $[2.01,3.13]$\\
$70$ & Extinction threat & $1.25$ & $[0.78,1.70]$\\
$100$ & Extinction probable & $0.62$ & $[0.34,0.94]$\\
$180$ & Chicxulub event & $0.141$ & $[0.062,0.249]$\\
\hline
\end{tabular}
\end{center}
\vspace{0.5cm}
\end{table}

\subsection{\label{sec:Ages}On extreme human ages: supercentenarians}

Although the terms `life expectancy' and `lifespan' are frequently used interchangeably, they are nevertheless defined differently: if the length of time for which one member of a given species is alive (i.e., 
the temporal interval between the member's birth and death) is defined as lifetime, then life expectancy at birth is the \emph{mean} lifetime of a population, whereas lifespan refers to the \emph{maximal} 
observed lifetime in that population. When used for individual members of a species, the life expectancy of one member at a given age is obtained after subtracting the current age of that member from the life 
expectancy at birth. Another related term is `longevity', the ability of one member of a species to live beyond the life expectancy at birth for that species. Extreme human longevity attracted Gumbel's 
attention already in the 1930s \cite{Gumbel1937}.

The lifetime of one member of a species depends on a number of factors: on the genes of that member, on the environment wherein that member lives, on pure chance, and (for humans) on individual habits (social, 
cultural) \cite{DeBenedictis2006}. Provided that the first three factors were arranged in order of decreasing importance in Ref.~\cite{DeBenedictis2006}, then heredity obviously features as the most important, 
followed by what I would call `ecology-related' factors (i.e., air quality, water and ground contamination, etc.) and (for humans and, to a lesser extent, for domesticated animals) `infrastructure-related' 
factors (i.e., the availability of and easy access to a reliable healthcare system). It comes as no surprise that `pure chance' also enters one's lifetime, as it does all natural processes: for the sake of 
example (borrowed from Particle Physics), $50$~\% of a set of charged pions, created at $t=0$, are expected to decay before $t \approx 26$ ns (to be identified with the mean life of the charged pion). 
Focussing on humans, each individual's lifetime also depends on that person's habits, i.e., on general lifestyle, on the regularity (which appears to be more important than the intensity) of physical 
exercise, on personal hygiene, on proper nutrition, and so on; smoking and alcohol consumption are generally accepted as life-shortening factors. Last but not least, there are also the 
`employment/profession-related' factors in modern human societies (stress, sedentary occupation, exposure to harmful substances/materials, and the like). Many of the factors, categorised as `environmental' 
in Ref.~\cite{DeBenedictis2006}, improved over the course of the past one century, the result being a sizeable increase in the human life expectancy at birth after World War \RNum{2}.

Verified records of \emph{super}centenarians, i.e., of humans over 110 years old \cite{GRG}, are publicly available \cite{ListOfSupercentenarians}; this study uses part of the data relating to the $100$ oldest 
female and the $100$ oldest male supercentenarians. Although included in the two lists are also data corresponding to \emph{living} supercentenarians, these (seven) subjects will be excluded from the analysis.

After examining the two lists of Ref.~\cite{ListOfSupercentenarians}, one cannot but take notice of a number of perplexing issues. To start with, there are no records of supercentenarians in Africa. Moreover, 
there is no mention of any Chinese, Indian, and Russian subjects in the two lists. In all probability, the lack of subjects from African countries relates to reliability, given that the formal registration 
of new births was not mandatory in many African countries before the second half of the $20$-th century; this could also apply to India, which was under British rule before 1950. Given the near obsession of 
the Chinese with observations (I have their old astronomical observations in mind), the lack of Chinese subjects in the two lists is surprising. Of course, it may well be that the lack of Chinese and Russian 
subjects does not imply lack of \emph{availability} of records, but lack of \emph{verification} of the records by the Gerontology Research Group (GRG) \cite{GRG}, the non-profit organisation in charge of 
validating such data. My efforts to clarify this question did not bring fruit, as my e-mail (on 7 June 2024) to the director of the GRG, R.D.~Young, remains unanswered.

The two lists of this work are dominated by North Americans (76 subjects: 66 from the USA, 5 from Canada, and 5 from Mexico) and Japanese (47 subjects). There are 48 European subjects (11 from France, 9 from 
Italy, 8 from the UK, 7 from Spain, 5 from Portugal, 3 from the Netherlands, 2 from Poland, and 1 from each of the following countries: Germany, Norway, and Romania), 13 Central and South Americans (4 from 
Brazil, 3 from Colombia, and 1 from each of the following countries: Argentina, Chile, Costa Rica, Ecuador, El Salvador, and Peru), and 6 from the Caribbean (3 from Puerto Rico and 1 from each of the following 
island countries: Barbados(!), Cuba, and Jamaica). The lists are completed with one Australian and one Israeli.

After having established the DB of the supercentenarians, one can estimate the difference - in days (d) - between the day of birth and the day of death of each subject; or so it should be. The first attempt 
at using Microsoft Excel for this evaluation called the fact to my mind that the application does not recognise dates before 1900. As a result, I decided to create my own C++ code for the evaluation (it is a 
simple exercise) and use Excel to validate it (for birth dates after 1 January 1900). Curiously, the two sets of results differed in two cases. Examination of the results unearthed the reason for the observed 
discrepancies: the year 1900 is a leap year for Excel~\footnote{Two of the ten supercentenarians, with 1900 as the birth year, had been born in February.}! (After creating my own code, it dawned on me that 
the easiest way of solving the problem at hand would have been to just add 400 years to the years of birth and death of the subjects, and take the difference using Excel's function DAYS; Excel recognises the 
year 2300 as an ordinary year.)

\begin{figure}
\begin{center}
\includegraphics [width=15.5cm] {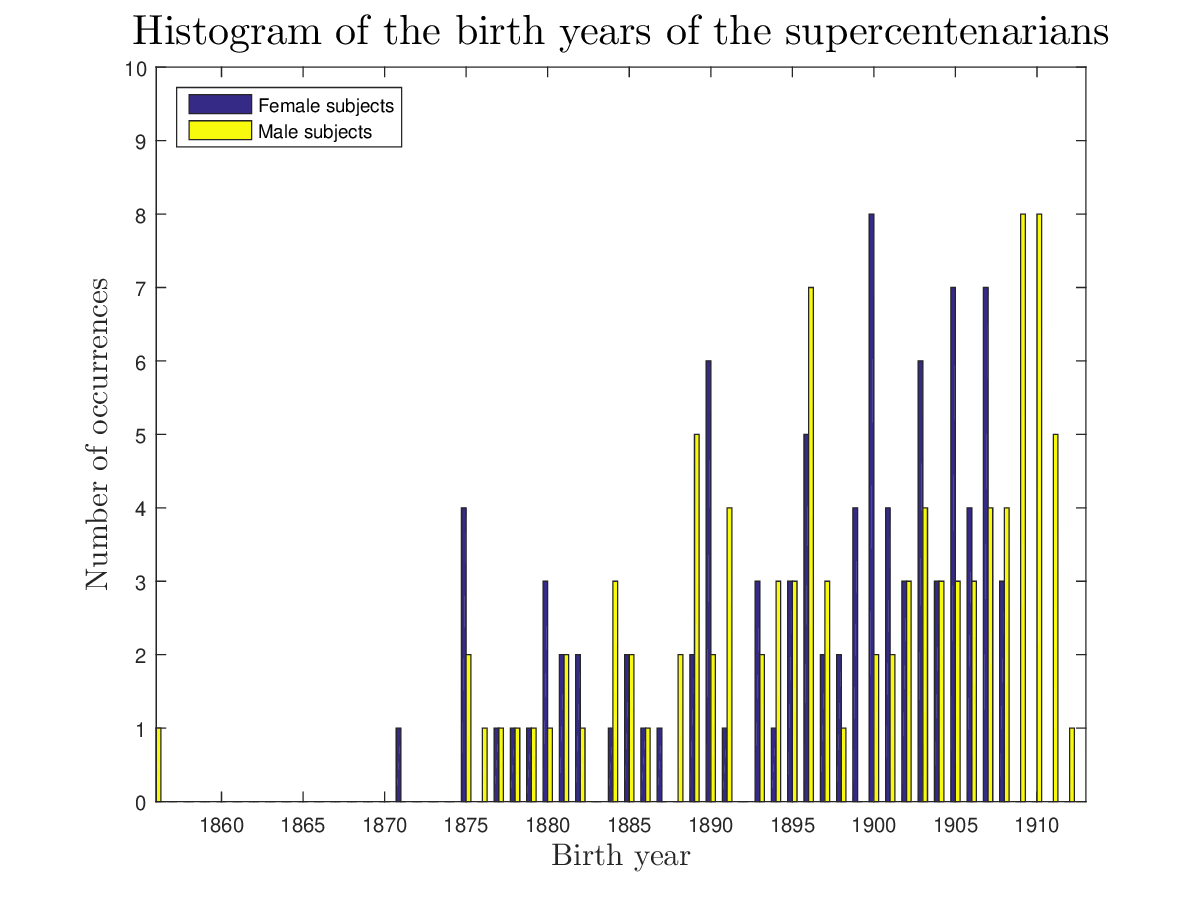}
\caption{\label{fig:HistogramAges}The two histograms of the birth year of the 94 female and the 99 male supercentenarians of the initial DB (containing supercentenarians with definite lifetime, with age 
determined at death) of this study.}
\vspace{0.5cm}
\end{center}
\end{figure}

The birth years of the female supercentenarians range between 1871 and 1908; there are no birth occurrences in the years 1872-1874, as well as in 1876, 1883, 1888, and 1892. The birth years of the male subjects 
range between 1856 and 1912; there are no birth occurrences in the years 1857-1874, as well as in 1883, 1887, 1892, and 1899. Given that there are no adjacent bins with zero occurrences between 1875 and 1908, 
the data of the female subjects may be analysed using a bin width of $2$ yr ($17$ final bins), hence $\Delta t = 2$ yr in this section. The same applies to the data of the male subjects between 1875 and 1912 
($19$ final bins). The earliest-born supercentenarian in either set (born in 1871 in the former set and in 1856 in the latter) needs to be removed from the initial DB. The two histograms of the birth year of 
the female and the male supercentenarians of the initial DB of this study are shown (in the form of two bar plots next to one another) in Fig.~\ref{fig:HistogramAges}. The ages of the female and the male 
supercentenarians of the analysis DB of this study were histogrammed in bins of the birth year, and the largest age was selected in each histogram bin: the selected values are given in Table \ref{tab:HistogramAges}, 
separately for the female and the male supercentenarians.

\vspace{0.5cm}
\begin{table}[h!]
{\bf \caption{\label{tab:HistogramAges}}}The ages of the female and the male supercentenarians of the analysis DB of this study were (separately) histogrammed in bins of the birth year and the largest age 
was selected in each histogram bin; this table contains the selected values. The earliest-born supercentenarian in either set (born in 1871 in the former case and in 1856 in the latter) was removed from the 
initial DB (see text and Fig.~\ref{fig:HistogramAges}).
\vspace{0.25cm}
\begin{center}
\begin{tabular}{|l|c|c|}
\hline
 & Female supercentenarians & Male supercentenarians\\
\hline
Birth years & Largest age (d) & Largest age (d)\\
\hline
\hline
$1875-1876$ & $44\,724$ & $41\,043$\\
$1877-1878$ & $42\,231$ & $41\,202$\\
$1879-1880$ & $43\,560$ & $41\,411$\\
$1881-1882$ & $42\,322$ & $42\,255$\\
$1883-1884$ & $42\,010$ & $41\,547$\\
$1885-1886$ & $42\,045$ & $41\,320$\\
$1887-1888$ & $42\,028$ & $41\,069$\\
$1889-1890$ & $42\,715$ & $41\,826$\\
$1891-1892$ & $42\,127$ & $42\,159$\\
$1893-1894$ & $42\,223$ & $41\,242$\\
$1895-1896$ & $42\,468$ & $41\,842$\\
$1897-1898$ & $42\,760$ & $42\,422$\\
$1899-1900$ & $42\,994$ & $41\,363$\\
$1901-1902$ & $42\,815$ & $41\,005$\\
$1903-1904$ & $43\,572$ & $41\,604$\\
$1905-1906$ & $42\,593$ & $41\,452$\\
$1907-1908$ & $42\,600$ & $41\,277$\\
$1909-1910$ & $-$ & $41\,949$\\
$1911-1912$ & $-$ & $41\,054$\\
\hline
\end{tabular}
\end{center}
\vspace{0.5cm}
\end{table}

\begin{figure}
\begin{center}
\includegraphics [width=15.5cm] {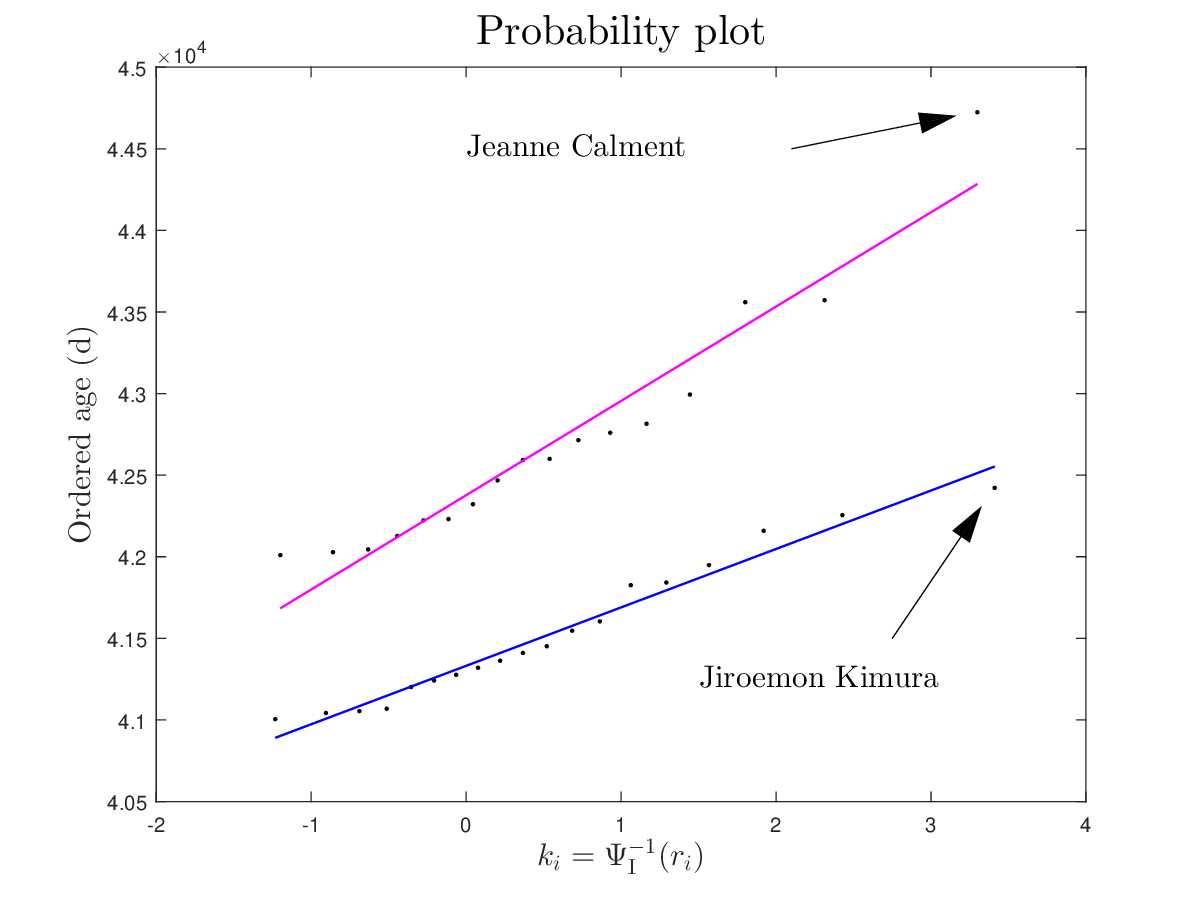}
\caption{\label{fig:ProbabilityPlotAges}The probability plot for the largest ages of the female and the male supercentenarians of Table \ref{tab:HistogramAges}. The values of the parameters $\mu$ and $\alpha$ 
for the female supercentenarians, obtained via ordinary linear regression (the fitted results of which are represented by the magenta solid straight line), are equal to $42\,377$ d and $578$ d, respectively. 
The corresponding values for the male supercentenarians are equal to $41\,332$ d and $358$ d; in this case, the fitted results of the ordinary linear regression are represented by the blue solid straight line.}
\vspace{0.5cm}
\end{center}
\end{figure}

The ordered set of the largest ages of Table \ref{tab:HistogramAges} are plotted against the uniform order statistic medians $k_i = \Psi^{-1}_{\rm \RNum{1}} (r_i)$ in Fig.~\ref{fig:ProbabilityPlotAges}; the 
plotting positions $r_i$ have been obtained from Eq.~(\ref{eq:EQ031_5}) with $A=3/8$, which (in comparison with $A=0$) yields a slightly larger sum of the PCC values for the two sets of data. The PCC between 
the arrays of the ordered ages and $k_i$ is about $0.9673$ for the female subjects and about $0.9860$ for the male subjects. Inspection of Fig.~\ref{fig:ProbabilityPlotAges} suggests that it will take longer 
to have the astonishing record set by Jeanne Calment~\footnote{The issue of Jeanne Calment's longevity has not been entirely devoid of controversy, e.g., see Ref.~\cite{Zak2019}.} broken; she outlived the 
second in rank, Kane Tanaka, by an impressive $1\,152$ d, which is over $3$ yr, a remarkable feat considering the advanced age. On the contrary, the age difference between the current record holder for male 
subjects, Jiroemon Kimura, and the second in rank, Christian Mortensen, is $167$ d, i.e., less than half a year. The standard deviation of the largest ages in the set of the male supercentenarians is 
considerably smaller than that of the female supercentenarians ($433$ versus $709$ d); however, this difference is mostly due to Jeanne Calment's outstanding longevity.

Table \ref{tab:ParametersAges} contains the results of the minimisation of the function MF of Eq.~(\ref{eq:EQ036}) for the parameters $\mu$ and $\alpha$ of Eq.~(\ref{eq:EQ017_5}): fitted values and 
uncertainties are given, separately for the female and the male supercentenarians of Table \ref{tab:HistogramAges}. The off-diagonal element of the Hessian matrix is equal to about $0.285613$ in the former 
case and about $0.308696$ in the latter.

\vspace{0.5cm}
\begin{table}[h!]
{\bf \caption{\label{tab:ParametersAges}}}The results of the optimisation to the two sets of the largest ages of Table \ref{tab:HistogramAges}. The t-multiplier, corresponding to $1 \sigma$ effects in the 
normal distribution, has been applied to the quoted uncertainties.
\vspace{0.25cm}
\begin{center}
\begin{tabular}{|l|c|c|c|c|}
\hline
 & \multicolumn{2}{|c|}{Female supercentenarians} & \multicolumn{2}{|c|}{Male supercentenarians}\\
\hline
Parameter & Fitted value & Corrected fitted & Fitted value & Corrected fitted\\
 & & uncertainty & & uncertainty\\
\hline
\hline
$\mu$ (d) & $42\,405$ & $117$ & $41\,333$ & $80$\\
$\alpha$ (d) & $446$ & $94$ & $323$ & $63$\\
\hline
\end{tabular}
\end{center}
\vspace{0.5cm}
\end{table}

\begin{figure}
\begin{center}
\includegraphics [width=15.5cm] {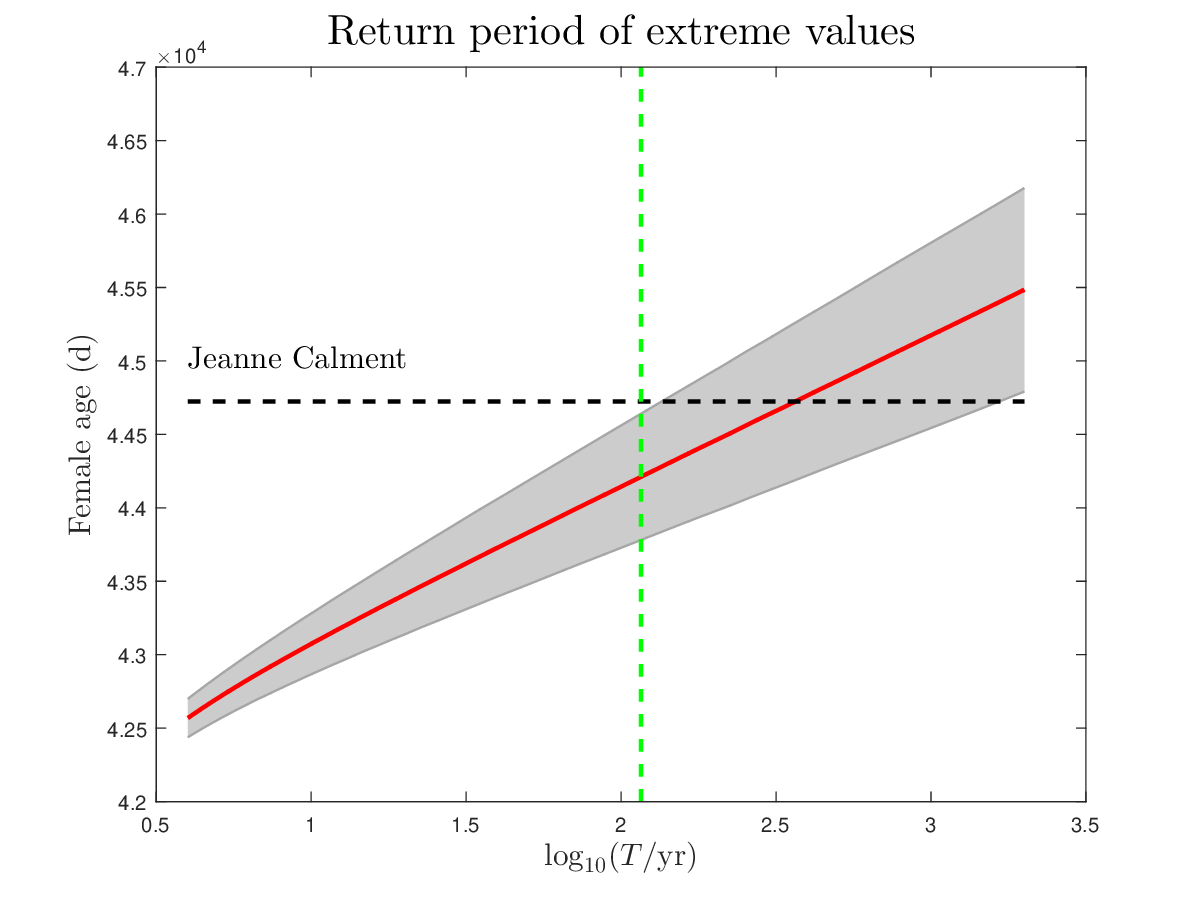}
\caption{\label{fig:ReturnPeriodAgesFemales}The relation between the expectation values of the future extreme age of the female supercentenarians and the return period. The red solid curve represents median 
values, whereas the two grey solid curves delimit the CIs which are associated with $1 \sigma$ effects in the normal distribution (i.e., with ${\rm CL} \approx 68.27$~\%). The black dashed horizontal straight 
line marks the age of the current record holder, Jeanne Calment. The green dashed vertical straight line marks the year 2024, corresponding to $116$ yr into the future of the historical measurements.}
\vspace{0.5cm}
\end{center}
\end{figure}

\begin{figure}
\begin{center}
\includegraphics [width=15.5cm] {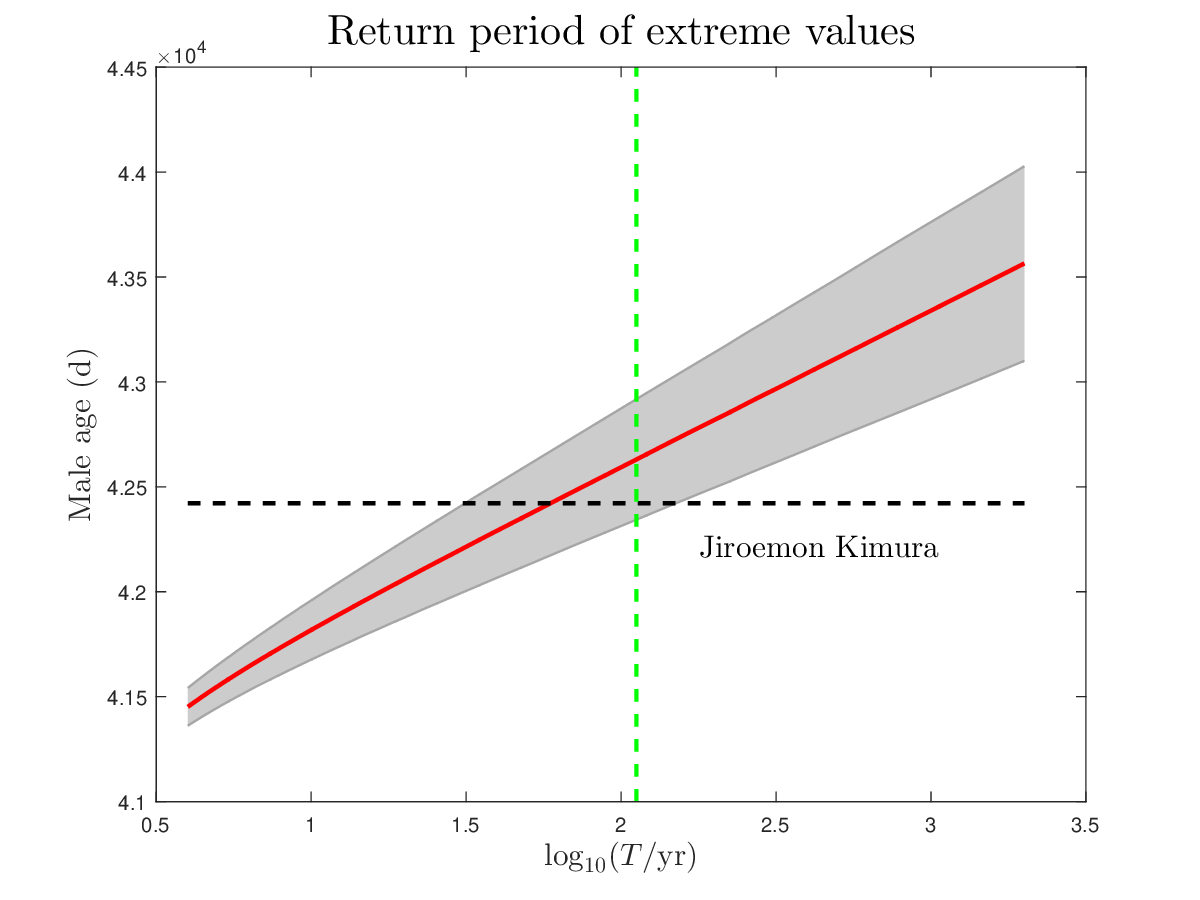}
\caption{\label{fig:ReturnPeriodAgesMales}The equivalent of Fig.~\ref{fig:ReturnPeriodAgesFemales} for the male supercentenarians. The black dashed horizontal straight line marks the age of the current record 
holder, Jiroemon Kimura. The green dashed vertical straight line marks the year 2024, corresponding to $112$ yr into the future of the historical measurements.}
\vspace{0.5cm}
\end{center}
\end{figure}

I shall finally address the question as to when the next two female and male record holders are expected to be born. As in Sections \ref{sec:Floods} and \ref{sec:Meteorites}, these predictions will be 
obtained via Monte-Carlo simulations using the results of the two optimisations to the data, i.e., the fitted parameter values and uncertainties of Table \ref{tab:ParametersAges}, along with the Hessian 
matrices of the two MLE fits. The results are shown in Figs.~\ref{fig:ReturnPeriodAgesFemales} and \ref{fig:ReturnPeriodAgesMales}; the comparison of these two figures leaves no doubt that, it will take 
longer to have the record of the age of the female supercentenarians broken; the new female record holder is expected to be born between the years 2043 and 3569 ($\approx 68.27$~\% CI), the expectation 
value of the median birth year being 2273. On the contrary, the new male record holder is expected to have already been born ($\approx 68.27$~\% CI of the birth year: 1943 to 2062), the expectation value 
of the median birth year having been 1971.

\section{\label{sec:Conclusions}Conclusions and discussion}

Presented in a concise form in this study were the essentials of the Extreme-Value Theory (EVT), which enables 
\begin{itemize}
\item the analysis of extreme values, obtained via the observation of extraordinary/rare physical phenomena, and
\item the extraction of reliable predictions relating to the future occurrence of such phenomena.
\end{itemize}
Emphasis has been placed on the application of the EVT to real-world data. As a result, the theoretical part of this paper includes those of the details which are necessary for the purposes of the analyses 
conducted herein, and omits others (e.g., the methodology to deal with statistical bias and departure from stationarity, issues which have been discussed in the literature and might be addressed in a future 
work).

The presentation of the formalism is tailored to the chronological order of the development of the EVT, starting with the early efforts by statisticians to invent a non-parametric, model-independent (i.e., 
distribution-free) method of analysis based on Order Statistics, see Section \ref{sec:DistributionNumberOfExceedances}. The theoretical framework of the EVT was established in the late 1920s, an important 
cornerstone being the 1928 paper by Fisher and Tippett \cite{Fisher1928}, who demonstrated that, regardless of the underlying distribution from which the data (yielding the extreme values) are drawn, the 
distribution of the extremes (as the sample size increases indefinitely) can only be one of three classes of functions: Gumbel, Fr{\'e}chet, or Weibull, see Section \ref{sec:MEVGeneralFormalism}. The second 
class had been discovered by Fr{\'e}chet \cite{Frechet1927} one year before Ref.~\cite{Fisher1928} appeared. The Gumbel class achieves the optimisation of the description of the input data by variation of 
two parameters, the two other classes by variation of three.

The general procedure, followed in modern extreme-value analyses, is detailed in Section \ref{sec:MEVAnalysis}. One usually starts with the probability plot, for the purposes of which the ordered input data 
are plotted against the `theoretical' quantiles, identified with the uniform order statistic medians $k_i \coloneqq \Psi^{-1} (r_i)$, where $r_i$ are the plotting positions of Eq.~(\ref{eq:EQ031_5}) and the 
quantile function $\Psi^{-1}$ refers to any of the three choices in Eqs.~(\ref{eq:EQ017}); the parameter $A \in [0,1]$ may be chosen in such a way that the Pearson Correlation Coefficient (PCC) between the 
plotted quantities be maximised. Although it is generally unclear which of the three classes of Eqs.~(\ref{eq:EQ017}) should be used in a specific extreme-value analysis, a choice may rely on the maximisation 
of the PCC in the probability plot~\footnote{The second option is to perform the fits to the extreme values using all three classes and select the class which provides the best description. The third option 
is to make use of the GEV distribution of Eq.~(\ref{eq:EQ018}) in the optimisation.}; this approach would lead to the fixation of the parameter $\zeta$ (entering the Fr{\'e}chet and Weibull classes), a 
welcoming prospect as this possibility reduces the correlations (among the model parameters) in the ensuing optimisation.

For the optimisation of the description of the input data, I recommend the use of the method of the Maximum-Likelihood Estimation (MLE). The maximisation of the likelihood is mathematically equivalent to the 
minimisation of the functions MF of Eqs.~(\ref{eq:EQ036},\ref{eq:EQ038},\ref{eq:EQ040}) for the three classes of functions. Provided that the parameter $\zeta$ (in case of the Fr{\'e}chet and Weibull classes) 
is fixed from the probability plot, all fits involve the variation of two parameters: the location parameter $\mu$ and the scale parameter $\alpha$, which associate the variable $x$ (extreme values) with the 
reduced variable $y$ of Eqs.~(\ref{eq:EQ017}), see Eq.~(\ref{eq:EQ017_5}). The output of each optimisation comprises: the fitted (optimal) values of the parameters $\mu$ and $\alpha$, their fitted 
uncertainties (corrected for the sample size via the application of the t-multipliers), along with the Hessian matrix of the fit.

Regarding the frequency of occurrence of extraordinary events, predictions can be obtained from Monte-Carlo simulations based on the aforementioned output of each optimisation: plots of the threshold $y_t$ 
of the extreme values against the corresponding return period of Eq.~(\ref{eq:EQ017_4}) provide answers to two types of questions, involving the return period of extraordinary events and their expected 
severity within a specific temporal window, see also caption of Fig.~\ref{fig:ReturnPeriodFloods}.

Three examples of the application of the EVT were examined in this study: Section \ref{sec:Applications} contains all results of the statistical analysis of the corresponding data.
\begin{itemize}
\item The first application involved the analysis of the very same data which Gumbel used in 1941, in the first ever application of the EVT to a real-world problem \cite{Gumbel1941}: in that study, the 
author determined the return period of floods and extracted predictions from the historical records of the largest daily discharges (i.e., of the volumetric flow rates) of the Rh{\^o}ne river in the Lyon 
area in each year between 1826 and 1936. The measurements, which Gumbel used in that paper, had been retrieved from the `Archives du d{\'e}partement du Rh{\^o}ne et de la m{\'e}tropole de Lyon' (`Archives 
d{\'e}partemental et m{\'e}tropolitaines', Lyon). First, the data were copied from the archives by Coutagne for the purposes of his 1938 paper \cite{Coutagne1938}. Subsequently, the data were either copied 
from Coutagne's paper by Gumbel or sent to him as private communication. This study suggests that, in copying the values from one document to another, mistakes might have been made: on account of the 
mismatch between the values given in Table \RNum{3} of Gumbel's paper and two sums (involving those values) quoted in that paper, such a possibility gains momentum. Owing to the difficulty of examining the 
original records, there is no way of explaining the apparent discrepancies at this time.
\item The second application related to the meteorite impacts on the Earth. Inspection of the data suggested that the majority of the available DBs contain unverified structures, i.e., craters whose 
characteristics do not conform with those from meteorite impacts. Due to this reason, the small, but verified, DB by the Impact Earth Group \cite{IEDB} was used herein, enhanced with the results corresponding 
to six craters (three small: with diameters below $300$ m; and three medium-sized: with diameters between $3$ and $8.5$ km), whose omission from the DB of the Impact Earth Group is puzzling (to me); luckily, 
due to the analysis procedure in this study, the treatment of these six structures (inclusion in or exclusion from the analysis DB) has no impact on the results. The analysis suggested that the 
frequently-voiced claim that Chicxulub-sized craters occur every $100$ Myr is incorrect: meteorite impacts, which are capable of creating crater diameters of the Chicxulub size, are several times less frequent.
\item The third application related to extreme human longevity. Inspection of the DB which has been verified by the Gerontology Research Group (GRG) \cite{GRG}, a body of experts who are in charge of 
validating such data, revealed that there are no supercentenarians from African countries, China, India, and Russia. This is surprising, in particular in the case of China, given the long experience of the 
Chinese for recording their observations. The analysis of the data suggests that the current record of lifespan in the female population will be broken long after a new record is set in the male population.
\end{itemize}

The results of this study were obtained nearly effortlessly after the software was developed and tested. There have been no occasions in which the maximisation of the likelihood functions posed any 
problem to the MINUIT software application. All runs terminated successfully, including (of course) the numerical evaluation of the Hessian matrices in all cases. Excel files with the input data to this 
study, the results of the optimisation, as well as with the predictions obtained thereof, are available upon request.

\begin{ack}
I am very grateful to Davide Sardella for suggesting to me the analysis of the data on the meteorite impacts on the Earth, for sending to me links to some relevant DBs, and for his information regarding 
their content.

I am indebted to Camille T{\^e}te-David (Biblioth{\`e}que de l'ESTP, Cachan) for her readiness to allow me to inspect Coutagne's 1938 paper \cite{Coutagne1938} in Cachan.

I take this opportunity to highlight the importance of preserving \emph{in digital form} the historical measurements of the largest daily discharges $x_i$ of the Rh{\^o}ne river between 1826 and 1936 
\cite{Coutagne1938}. For one thing, these $111$ values were used in the first ever application of the EVT to a real-world problem. For another, this dataset contains measurements which are nearly devoid of 
direct systematic effects, i.e., of effects which - due to the rising global average temperature over the course of the past few decades - surely affect later data, see Refs.~\cite{Matsinos2023b,Matsinos2024} 
and the papers cited therein. All things considered, this dataset represents a textbook case of the application of the EVT. It is therefore important to keep a safe copy of these data. Although \emph{photocopying} 
material of old books might be harmful (to those books), \emph{photographing} a few important pages of a book and making the original material publicly available will be beneficial to the future generations 
of scientists and historians; to mention one notable example, digital photographs of the pages of Galton's notebook, containing his famous $19$-th century family data on human stature, were taken in 2003, 
see Ref.~\cite{Matsinos2023a} for details. I find it somewhat surprising that a similar effort to secure Coutagne's data has not yet been undertaken.

Figure \ref{fig:Sets} has been created with CaRMetal \cite{CaRMetal}. The remaining figures of this study have been created with MATLAB$\textsuperscript{\textregistered}$ (The MathWorks, Inc., Natick, 
Massachusetts, United States).
\end{ack}

\clearpage
\newpage
\appendix
\section{\label{App:AppA}Distribution of the largest value of normally-distributed data}

On p.~7-12 of his book \cite{Kinnison1983}, Kinnison addresses the distribution of the largest value of normally-distributed data. Unfortunately, the formulae, given on that page of the book without proof 
or reference to earlier work, contain mistakes: for instance, both expressions for his scale parameter a$(n)$ are wrong. In all probability, Kinnison copied the formulae from an earlier source, presumably 
from the 1946 book `Mathematical Methods of Statistics' by Harald Cram{\'e}r (1893-1985) \cite{Cramer1946}, but did not check whether the formulae he gives on that page make any sense.

The correct formulae can be obtained from Eqs.~(28.6.13,28.6.14) of Cram{\'e}r's book, p.~375. Provided that $x_i$ is a sequence of independent, identically-distributed normal random variables with average 
$m$ and standard deviation $\sigma$, then the CDF of the largest of $n$ such values, denoted herein as $M(n)$, belongs to the Gumbel class.
\begin{equation} \label{eq:EQA01}
P(M(n) \leq x) = \Psi_{\rm \RNum{1}}(y) = e^{-e^{-y}} \, \, \, ,
\end{equation}
where
\begin{equation} \label{eq:EQA02}
y = \frac{x - \mu(n)}{\alpha(n)} \, \, \, ,
\end{equation}
with
\begin{align} \label{eq:EQA03}
\mu(n) &= m + \sigma \, c(n),\nonumber\\
\alpha(n) &= \frac{\sigma}{\sqrt{2 \ln n}}, \text{and}\nonumber\\
c(n) &= \sqrt{2 \ln n} \left( 1 - \frac{\ln \left( 4 \pi \ln n \right)}{4 \ln n} \right) \, \, \, .
\end{align}

The quantity $y$ of this study is identified with the quantity $v$ of Cram{\'e}r's book. The PDF of Cram{\'e}r's Eq.~(28.6.14), $j_{\nu} (v)$, for $\nu=1$ (i.e., for the largest value of normally-distributed 
data), is identified with the PDF $\psi_{\rm \RNum{1}}(y)$ of this work.

\clearpage
\newpage
\section{\label{App:AppB}Extremes of small samples}

In Chapter 10 of his book \cite{Kinnison1983}, Kinnison deals with the subject of small reference datasets, mentioning in passing that ``there are some situations in which the asymptotic distributions [i.e., 
the expressions on the right-hand side of Eqs.~(\ref{eq:EQ017})] yield significantly biased results.'' However, no example of such a ``situation'' is given in the book. To remedy this probl{\'e}matique, 
Kinnison retraces his steps back to Order Statistics, yet without providing an explanation why he expects Order Statistics to be reliable in those ``situations.'' For several reasons, I find that Chapter 10 
of Kinnison's book seriously lacks in clarity and coherence; this may explain why I decided to provide a summary of the (few) essential points, which Kinnison strives to make in that part of his book. In 
fact, several of the author's sentences in that chapter became obsolete a few years after the book appeared.

In essence, Kinnison suggests that any predictions extracted from small reference datasets should not be based on the procedure outlined in Section \ref{sec:MEVAnalysis} of this work, but on a simpler method 
making use only of the rank of the input data. It is unclear to me why Kinnison expects that the use of the rank is more reliable than the procedure of Section \ref{sec:MEVAnalysis}. Be that as it may, two 
analysis procedures are better than none (or, given that one can compare the results, even one).

Let the (small) reference dataset contain $n$ values, drawn from an arbitrary distribution (whose CDF will be denoted by $F$) and arranged in ascending order. The probability that exactly $j$ out of the $n$ 
measurements will be smaller than or equal to a specific value $X$ is given by the expression
\begin{equation} \label{eq:EQB01}
C(n,j) \left( F(X) \right)^j \left( 1 - F(X) \right)^{n-j} \, \, \, ,
\end{equation}
where the binomial coefficient $C(a,b)$ has been defined in Eq.~(\ref{eq:EQ002}). The probability that at least $r$ values will be smaller than or equal to $X$ is thus given by the formula:
\begin{equation} \label{eq:EQB02}
P(x_r \leq X) = \sum_{j=r}^n C(n,j) \left( F(X) \right)^j \left( 1 - F(X) \right)^{n-j} \, \, \, .
\end{equation}
In spite of the plethora of formulae in Kinnison's Chapter 10, this is the only useful formula in that chapter (apart from the relations which pertain to the Bonferroni correction, a subject which I would 
include in another chapter of the book, probably in the form of an appendix). All approximations, mentioned in Chapter 10 of Kinnison's book, albeit of some historical significance in the era of pocket 
calculators and tabulated values of functions, are - by today's computational standards - redundant and (as they distract the reader's mind into directions of debatable practicality) misleading~\footnote{This 
remark should be taken as a fact, not as criticism.}.

From Eq.~(\ref{eq:EQB02}), it follows that the distribution of the smallest value will be
\begin{equation} \label{eq:EQB03}
P(x_1 \leq X) = \sum_{j=1}^n C(n,j) \left( F(X) \right)^j \left( 1 - F(X) \right)^{n-j} = 1 - \left( 1 - F(X) \right)^{n} \, \, \, ,
\end{equation}
whereas the distribution of the largest will be
\begin{equation} \label{eq:EQB03}
P(x_n \leq X) = \left( F(X) \right)^n \, \, \, .
\end{equation}

For the sake of example, Kinnison derives the upper $5$~\% limit of the $4$-th order statistic (second largest, given the order in which the input data are arranged) in case of a sample of dimension $5$; 
evidently, $r=4$, $n=5$, and $P(x_4 \leq X) = 1 - 0.05 = 0.95$. Solving Eq.~(\ref{eq:EQB02}) for its only unknown, $F(X)$, yields $F(X) \approx 0.9236$. The return period of events $x>X$ (which is equivalent 
to $x \geq X$ for continuous stochastic variables) is thus equal to $(1-F(X))^{-1} \, \Delta t \approx 13.1 \, \Delta t$.

\end{document}